\documentclass[journal]{IEEEtran}

\usepackage{graphicx,amssymb,amsmath,amsfonts}
\usepackage{array}
\usepackage[caption=false,font=normalsize,labelfont=footnotesize,textfont=footnotesize]{subfig}
\usepackage{textcomp}
\usepackage{stfloats}
\usepackage{url}
\usepackage{verbatim}
\usepackage{graphicx}
\usepackage{cite}
\usepackage[ruled,vlined]{algorithm2e}
\usepackage{xcolor}
\usepackage{siunitx}
\usepackage{tabularx} 
\usepackage{balance}
\usepackage{hyperref}
\usepackage{epstopdf}

\newcommand{\norm}[1]{\left\lVert#1\right\rVert}
\DeclareMathOperator{\taninv}{tan^{-1}}
\DeclareMathOperator*{\argmax}{arg\,max}

\begin{document}

\title{An Efficient Approach with Dynamic Multi-Swarm of UAVs for Forest Firefighting}
\author{Josy John, K. Harikumar, J. Senthilnath,~\IEEEmembership{Senior Member,~IEEE}, Suresh Sundaram,~\IEEEmembership{Senior Member,~IEEE}
        % <-this % stops a space
\thanks{Josy John and Suresh Sundaram are with the Department of Aerospace Engineering, Indian Institute of Science, Bengaluru, India. \href{mailto:josyjohn@iisc.ac.in}{josyjohn@iisc.ac.in};\href{mailto:vssuresh@iisc.ac.in}{vssuresh@iisc.ac.in}}% <-this % stops a space
\thanks{K. Harikumar is with the Robotics Research Center, International Institute of Information Technology, Hyderabad, India, \href{mailto:harikumar.k@iiit.ac.in}{harikumar.k@iiit.ac.in}}% <-this % stops a space
\thanks{J. Senthilnath is with the Department of Machine Intellection, Institute for Infocomm Research,
Agency for Science, Technology and Research (A*STAR), Singapore,  \href{mailto:senthil@i2r.a-star.edu.sg}{senthil@i2r.a-star.edu.sg}}}



\maketitle

\begin{abstract}
In this paper, the Multi-Swarm Cooperative Information-driven search and Divide and Conquer mitigation control (MSCIDC) approach is proposed for faster detection and mitigation of forest fire by reducing the loss of biodiversity, nutrients, soil moisture, and other intangible benefits. A swarm is a cooperative group of Unmanned Aerial Vehicles (UAVs) that fly together to search and quench the fire effectively. The multi-swarm cooperative information-driven search uses a multi-level search comprising cooperative information-driven exploration and exploitation for quick/accurate detection of fire location. The search level is selected based on the thermal sensor information about the potential fire area. The dynamicity of swarms, aided by global regulative repulsion and merging between swarms, reduces the detection and mitigation time compared to the existing methods. The local attraction among the members of the swarm helps the non-detector members to reach the fire location faster, and divide-and-conquer mitigation control ensures a non-overlapping fire sector allocation for all members quenching the fire. The performance of MSCIDC has been compared with different multi-UAV methods using a simulated environment of pine forest. The performance clearly shows that MSCIDC mitigates fire much faster than the multi-UAV methods. The Monte-Carlo simulation results indicate that the proposed method reduces the average forest area burnt by $65\%$ and mission time by $60\%$ compared to the best result case of the multi-UAV approaches, guaranteeing a faster and successful mission.
\end{abstract}


\section{Introduction}
\IEEEPARstart{F}{orest} fires are highly devastating, leading to massive loss of wildlife, vegetation, wildlife habitats, etc. Forest fires cause the menace of extinction of plants and animals, disturbing the biodiversity and the ecosystem. Unfortunately, the annual global destruction due to forest fire aggregates to several million hectares of forest lands. The detection and mitigation of nascent forest fires are imperative to curtail the damage. Forest survey of India issues alerts of forest fires based on real-time detection of forest fires by Moderate Resolution Imaging Spectro-radiometer (MODIS) and Visible Infrared Imaging Radiometer Suite (VIIRS) data. The number of forest fire detected in India during the 2019-2020 fire season records $22447$ and $124473$ alerts using MODIS and VIIRS sensors \cite{moef}. In India, total forest area affected by fire is estimated as $1.1094$ Mha, with a $95\%$ confidence level for the $2019-2020$ \cite{fsi}. The forest fire in the Bandipur Tiger Reserve and National Park in 2019 destroyed nearly $40,000$ acres of forestland. The traditional methods emphasize preventing forest fire by improving the resilience of the forests against fire hazards with several forest management policies. The traditional forest fire suppression methods constitute firefighting techniques like directly beating out the fire with hand tools, fuel separation in advance to active fire, etc. \cite{ndma}. The existing forest fire monitoring, detection, and mitigation techniques engaging human firefighters are dangerous and unreliable.

UAV systems have been used for surveillance, monitoring, detection, and mitigation of forest fires inorder to reduce human interaction in unsafe circumstances \cite{yuan2015survey}. The single UAV systems are inadequate for large and multiple wildfire suppression measures. Thus team-based multi-UAV systems are promising for multi-task missions and increased area coverage. Multi-task missions like forest firefighting include search, detection, verification, monitoring, and neutralization of the targets \cite{akhloufi2020unmanned}. Multi-UAV performing multi-task missions need coordination strategies to manage communication and task allocation. The centralized, distributed, and decentralized control systems with global and local communication have been used in different fire assistance systems \cite{yuan2015survey}, \cite{akhloufi2020unmanned}. In the case of forest firefighting, the unknown targets are located in an extensive search area. In a practical scenario, forest fire spread can be uneven and time-varying due to the direction and speed of the wind, vegetation type, other environmental conditions, land cover type, etc. The large search area, together with uncertain and challenging conditions, makes it difficult for the UAVs to manage centralized or distributed communication. A decentralized control with local communication between UAVs is preferred for achieving scalability and robustness in forest fire search missions \cite{senanayake2016search}.

In existing literature, fire monitoring problem is the most addressed research direction in firefighting. The satisfactory tracking of perimeter depends on the fire models used to estimate the fire front and fire behavior. The simplest models use an elliptical fire profile with a constant rate of increase for the semi-minor and semi-major axis. The rate of increase of fire depends on the available fuel energy, intensity of the fire, direction, and speed of wind \cite{byram1959combustion}, \cite{anderson1982modelling}. Multi-UAV systems are effective for monitoring the dynamic fire perimeter to provide real-time data to firefighters \cite{casbeer2005}. Recently formation control based approaches \cite{ghamry2016cooperative}, \cite{yu2021fault}, re-configurable approaches \cite{casbeer2006cooperative}, \cite{ghamry2016fault}, auction-based approaches \cite{sujit2007fire}, potential field-based algorithms \cite{Pham}, and meta-heuristic based approaches \cite{Ruiz2022real} have been used to monitor the fire front given the location. The existing monitoring methods require a priori information on the number and location of forest fire for planning the search and coverage path. The UAVs for the mision are launched closer to the fire location, assuming that all vehicles are within a specific range of each other and the base station for uninterrupted communication.

The search task is one of crucial tasks in forest firefighting, considering the time required to locate the spreading fire in a large unknown area. Different deterministic and stochastic strategies have been proposed for multi-UAV search and destroy missions \cite{george2011search}. Stochastic search methods are efficient for missions involving large unknown search areas with limited or no communication, and Markovian search strategies are more efficient for dynamic targets \cite{flenner2012levy}. The optimal animal foraging behavior is well modeled by random walks, in which the predator optimizes the search for prey or food. Brownian and levy flight are the major random walk strategies found in searching behavior of different classes of animals including marine predators \cite{viswanathan2000levy}, \cite{bartumeus2003helical}. A multi-level search strategy inspired from the foraging behavior of the marine predator, Oxyrrhis Marina is used for a multi-UAV mission to search the forest fire locations \cite{OMS}. A combination of Levy, Brownian and directionally driven Brownian search is used to detect the fire locations. A two-step search and monitoring procedure for forest fire with a group of UAVs has been addressed in \cite{SARKAR2021}. The grid-based sandclock pattern search developed has a shorter path for scan within the grid cell. The monitoring strategy discovers the boundary points of the fire for the fire area estimation considering an arbitrary fire spread model and safe hovering time for the UAVs above the fire zone.

The monitoring and mitigation of forest fire with multiple UAVs is first addressed in \cite{Kumar2011}. An artificial potential field-based control law coordinates the motion of UAVs to track and quench the fire, avoiding the collision. The UAVs distributed uniformly over the fire boundary extinguish a circular area of the fire directly beneath the UAVs. The efficiency of fire extinguishing balls for firefighting is experimentally verified in \cite{aydin2019use} and results indicate that fire extinguishing balls might be effective in extinguishing short grass fires if a swarm of UAVs drop at an optimal location in optimal numbers. A conceptual framework for UAV assisted fire mitigation is explained in \cite{ausonio2021drone}. The system involves a platform for the UAVs to manage the replacement of batteries and payloads. The linear meters of fire front that can be extinguished is calculated based on the payload capacity of the UAV, capacity of the platform, time for reaching the fire front, and critical flow rate. In \cite{OMS}, UAVs use the Dynamic Formation Control (DFC) to mitigate the fire by spraying water covering an equal non-overlapping fire area between all UAVs acting on the same fire. The cumulative area quenched by all the UAVs should be less than the active fire area to maintain a non-overlapping area. In \cite{OMS}, UAVs mitigate overlapped areas when the area shrinks below a value and the overlapped area needs to be accounted as a buffer time in total mission time. Even though the methods above cover search, detection, and mitigation of forest fire, the distributed multi-UAV methods have longer detection and mitigation time as each UAV acts independently. The cooperative swarm-based search and mitigation of forest fires results in faster detection and mitigation as the members of the swarm cooperate and act together during the mission.

% The control law efficiently tracks the desired angular displacement between neighboring UAVs to cover an equal non-overlapping fire area between all UAVs acting on the same fire.The multi-UAV system is a possible solution for the problems involving large search areas and unknown dynamic targets, like forest fire. However, forest fires need simultaneous and continuous attention for efficient mitigation. The existing multi-UAV methods are inefficient in forest fire mitigation unless a sufficiently large number of UAVs are involved as each UAV is searching and mitigating independently.

In this paper, a Multi-Swarm Cooperative Information-driven search and Divide and Conquer mitigation control (MSCIDC) is presented for search and mitigation of forest fire. Multi-swarm missions are efficient in quenching the spreading fires as there is a high likelihood of multiple UAV detecting the same target. Multi-swarm cooperative information-driven search is used in the target search phase to drive the swarm in the direction of maximum sensor information. The search phase uses cooperative information-driven exploration and exploitation depending on the sensor information of members in the swarm. The cooperative nature between swarm members helps it to move toward the maximum temperature gradient sensed by the swarm. The swarm member detecting the fire locally attracts other members in that group to reach the fire front faster. The swarm members reaching the fire front start quenching fire using divide and conquer mitigation control. The quenching members are assigned to non-overlapping sectors of fire, and the sector division assures a non-overlapped quenching even with a shrinking fire area. The cooperative swarms with global regulative repulsion and merging, balance target detection and quench time reduction. The MSCIDC is assessed for a forest fire scenario in the pine forest, and the performance is compared with existing multi-UAV search methods. The performance of MSCIDC is also analyzed for varying numbers of swarms with the same number of UAVs. The Monte-Carlo simulations are carried out to verify the results. The results of MSCIDC are superior to the best case of multi-UAV missions with a $60\%$ reduction in total mission time and $65\%$ reduction in burnt area of the forest. 

The rest of this article is organized as follows. In Section II, problem definition and different models of subsystems are presented. Section III explains the information-driven multi-swarm search and subsequently divide and conquer mitigation control. Numerical simulation results are given in Section V to demonstrate the performance of the proposed method over the multi-UAV methods. Section VI concludes the paper.

\section{Problem Definition and Modelling}
\begin{figure}[htbp]
	\centerline{\includegraphics[width=75mm]{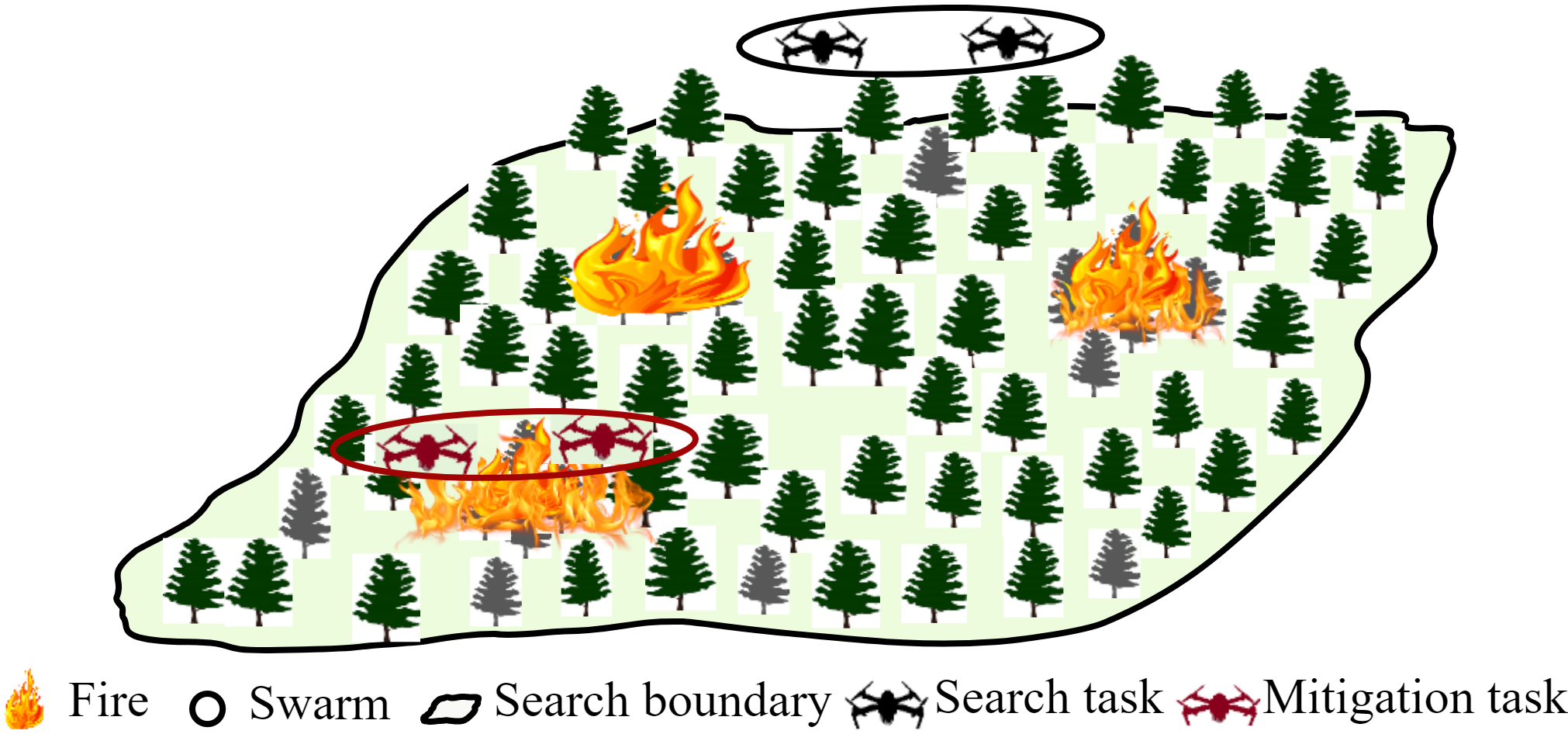}}
	\caption{A forest firefighting scenario where multiple swarms of UAVs perform search and mitigation of forestfire.}
	\label{swarmsearchschematic}
	\vspace{-5.5mm}
\end{figure}

The primary goal of forest firefighting is to prevent biodiversity loss by detecting and mitigating fires faster. The forest fire spreads faster based on the available biomass fuel, wind, and terrain conditions. The fire regions are often difficult to access from the ground when the forest fire spreads. UAV missions are more reliable for for detecting and neutralizing multiple fire points located in vast inaccessible forest regions. Multiple UAVs should act simultaneously and continuously for efficient mitigation of the fast-spreading forest fires. A typical forest fire scenario in a pine forest is shown in Fig. \ref{swarmsearchschematic}. Multiple swarms of UAVs are employed for the detection and mitigation of multiple unknown fire locations. The swarm marked in black is searching the area, and the other swarm marked in red is quenching a fire area by flying along the fire front. The cooperative behavior of members of the swarm leads to faster detection and mitigation of targets.\\
Let the area to search be $\Omega\subset {\rm I\!R}^2$, and the boundary of the search area is marked in black color, as shown in Fig. \ref{swarmsearchschematic}. Here, multiple swarms of UAVs search the area for multiple forest fire locations. Let $n_s$ be the number of swarms, and the search area is known to all the swarms. Let $S =s_1,s_2....s_{n_s}$ be the set of multiple swarms of UAVs deployed in the search area to detect and mitigate fire locations. Let swarm $s_i\in S$ has $n_{s_i}$ number of UAVs where $i=1,2,.....,n_s$. Let $n_f$ be the number of unknown fire locations within the search area. The fire profile is approximated as a circle or ellipse in this work. The coordinates of the $j^\text{th}$ fire profile are denoted as $p_{f_j}\in {\rm I\!R}^2$ with fire center locations $C_{f_j}\in {\rm I\!R}^2$ where $j=1,2,.....,n_f$. Let the major and minor axis lengths of $j^\text{th}$ fire be $a_j(t)$ and $b_j(t)$, respectively. The location and spread area of fire points are unknown to the searching agents. The fire spread rate depends on the affected area's topography, fuel content, terrain, wind speed, etc. It is necessary to detect the fire at a nascent stage to reduce the quenching time and destruction of biodiversity. The objective of the multi-swarm mission is to detect the maximum number of fire locations and quench the fire in the shortest duration possible. The objective is mathematically expressed using different models of associated subsystems as given in the following subsections.
\subsection{Fire Spread Model}
The circular or elliptical approximation is the most commonly used fire spread model. The fire size grows depending on the rate of fire spread, and the dynamic spread of the fire can be expressed as a function of fireline intensity, $I_l$ \cite{hansen2012},\cite{byram1959combustion}. The fireline intensity is defined as the heat release rate per unit length of the fire front, and $I_l$ in (\si{kW/ m}) is given by
\begin{equation}
	I_l = \alpha*(L_f)^\beta
\end{equation}
where $L_f$ is the flame length in \si{m}, $\alpha$ and $\beta$ are coefficients that depends on the available fuel content. The fire spread rate ($R$) in \si{(m/s)} is given by
\begin{equation}
	R=\frac{I_l}{H_cF_m}
\end{equation}
where $H_c$ is the heat of combustion in \si{kJ/ kg} and $F_m$ is the mass of the available fuel per unit area in \si{kg/m^2}. The fire spread rate is assumed to be constant and is calculated assuming a uniform fuel availability, terrain, and wind conditions in the search area. For a typical pine forest scenario, the values $\alpha=$ 259.833, $\beta=$ 2.174, $L_f=$ 4 \si{m}, $H_c=$ 18600 \si{kJ/ kg} and $F_m=$ 4 \si{kg/m^2} are considered \cite{hansen2012},\cite{penney2019calculation}.
\subsection{Swarm Model}
\begin{figure}[htbp]
	\centerline{\includegraphics[width=65mm]{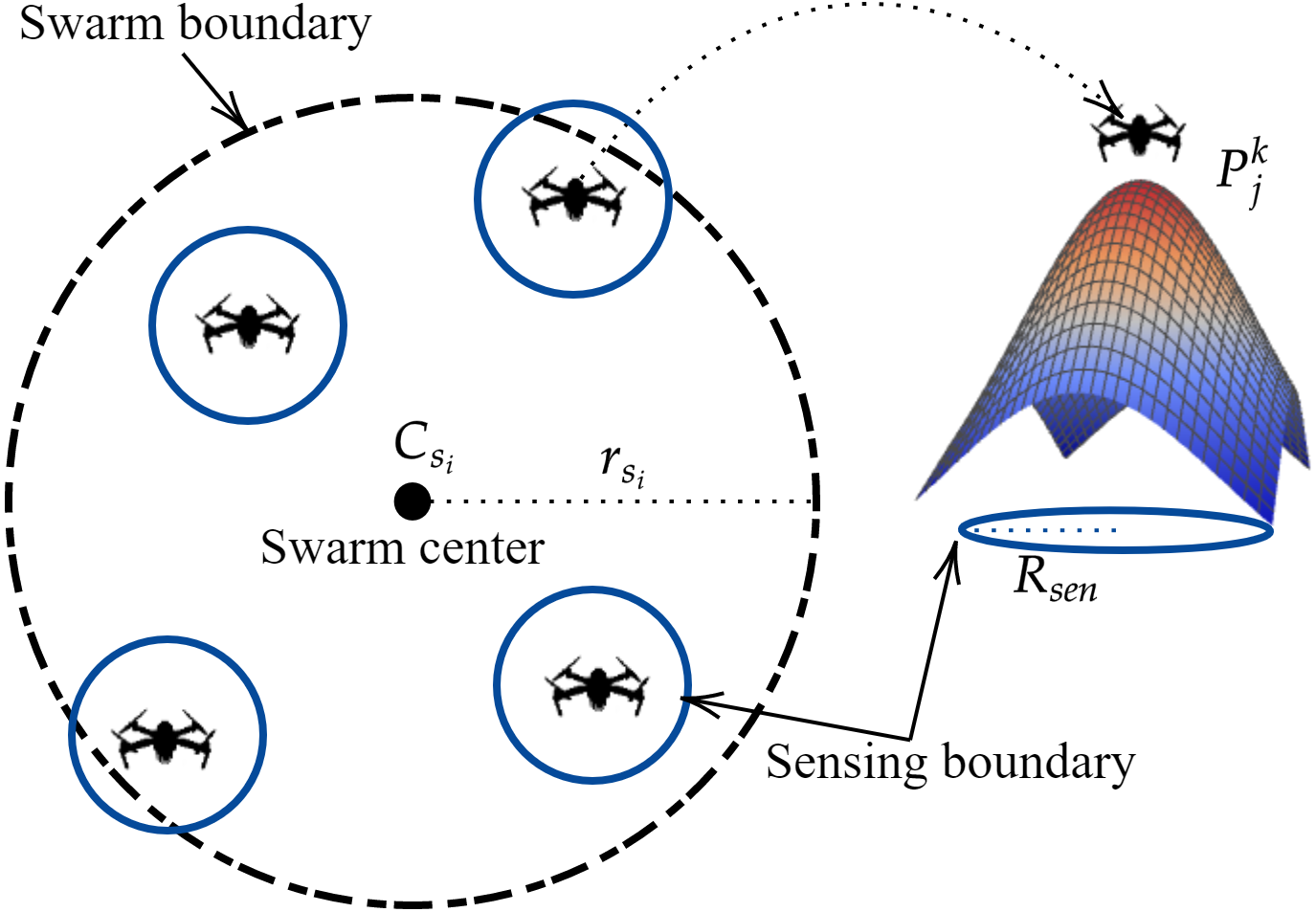}}
	\caption{Illustration of the swarm model showing sensing boundaries.}
	\label{swarmpic}
\end{figure}

A swarm is a set of UAVs flying together cooperatively, keeping the swarm intact. The swarm members are free to move independently within the swarm boundary. UAVs in each swarm have an attractive force to the center of the swarm to keep the swarm intact. The schematic diagram of a typical swarm with four members is shown in Fig. \ref{swarmpic}.
A circular swarm boundary with a constant swarm radius, $r_{s_i}$ is assumed for all the swarms. The swarm center, $C_{s_i}\in {\rm I\!R}^2$ is defined as the mean of the positions of all UAVs in the $s_i^\text{th}$ swarm.
\begin{equation}
	C_{s_i}=\frac{1}{n_{s_i}}\sum_{k=1}^{n_{s_i}} p_{s_i}^{\,k}
\end{equation}	
where $p_{s_i}^{\,k}=\begin{bmatrix}
	x_{s_i}^{\,k}&y_{s_i}^{\,k}
\end{bmatrix}$ is the position of $k^\text{th}$ swarm member of the $s_i^\text{th}$ swarm. Each member of a swarm satisfies the condition given below.
\begin{equation}
	\norm{p_{s_i}^{\,k}-{C_{s_i}}}<r_{s_i}
\end{equation}
% The distance between the swarm member to the swarm center of the corresponding swarm should always be within the defined swarm boundary.
% If $\norm{p_{s_i}^{\,k}-{C_{s_i}}}\ge r_{s_i}$, the attractive force of swarm center pulls back the swarm member towards the swarm center in the same direction. 
\subsection{UAV Kinematic Model}	
Let $v_{s_i}^{\,k}\in {\rm I\!R}^2$ and $v_{rs_i}^{\,k}\in {\rm I\!R}^2$ be the velocity and reference velocity vector of $k^\text{th}$ swarm member of the $s_i^\text{th}$ swarm at a time, $t$. All UAVs are homogeneous and belong to multi-rotor category. The mathematical model governing the motion of UAV is given by a first-order system.
\begin{flalign}
	\nonumber
	&\dot{{p}_{s_i}^{\,k}}(t)=v_{s_i}^{\,k}(t)\\
	&\dot{{v}_{s_i}^{\,k}}(t)=-\lambda_{s_i}^{\,k}v_{s_i}^{\,k}(t)+\lambda_{s_i}^{\,k}v_{rs_i}^{\,k}(t)
\end{flalign}
where $\lambda_{s_i}^{\,k}$ is the pole of first-order approximation of the mathematical model of the UAV. The reference velocity inputs for the position to velocity reference feedback are given by
\begin{flalign}
	&v_{rs_i}^{\,k}(t)=\frac{V_0(p_{rs_i}^{\,k}(t)-p_{s_i}^{\,k}(t))}{\tau+\|e_{s_i}^{\,k}(t)\|_2}+\dot{{p}_{rs_i}^{\,k}}\\
	&e_{s_i}^{\,k}(t)=p_{rs_i}^{\,k}(t)-p_{s_i}^{\,k}(t)
\end{flalign}
where $p_{rs_i}^{\,k}(t)$ and $\dot{{p}_{rs_i}^{\,k}}$ are the position and velocity of reference trajectory. $V_0 > 0$ is the cruise speed in \si{m/s}, $\tau>0$ is a small positive quantity, and $e_{s_i}^{\,k}(t)$ is the tracking error vector.
\subsection{Fire Detection Model}
A temperature detector and a thermal imaging sensor are mounted in each UAV for the detection of fire locations. The temperature detector measures the approximate temperature and rate of change of temperature.. The thermal imaging sensor gives the accurate fire profile and location. Fig. \ref{swarmpic} also shows the probability distribution as a function of the distance between UAV and fire location. Let $T_{s_i}^{\,k}$ be the temperature sensed by $k^\text{th}$ UAV, and $R_{sen}$ be the sensing radius within which fire detection is possible. The potential fire area identified by the UAV is confirmed using a Gaussian probability model. 
\begin{equation}
	P_j^{\,k} = exp(-\frac{d_{jk}^2}{2\sigma^2})
	\label{probconf}
\end{equation}
where $P_j^{\,k}$ is the probability of detection of the $j^\text{th}$ fire front by the $k^\text{th}$ UAV, $d_{jk}=\norm{p_{s_i}^{\,k}-p_{f_j}}$ is the distance between $k^\text{th}$ UAV and $j^\text{th}$ fire front, and $\sigma$ is the standard deviation of the sensor range. The fire location is detected if the probability of detection exceeds the detection threshold ($\gamma$). After the detection of fire, the swarm members can sense the fire profile data and start quenching the fire, which follows the quench model.
\subsection{Quench Model}
The critical flow rate ($CF$) is the water flow rate required to extinguish the fire with infinite time. UAVs can quench the fire area by spraying water at a higher rate than $CF$. The critical flow rate is proportional to the flame length, and $CF$ in \si{kg/m^{2}s} is given by
\begin{equation}
	CF=c*L_f^{\nu}
\end{equation}
where $c$ and $\nu$ are constants.

%\begin{figure}[htbp]
%	\centerline{\includegraphics[width=70mm]{newquenchvsarea.eps}}
%	\caption{Quench time for different number of UAVs.}
%	\label{quenchvsarea}
%\end{figure}
\begin{figure}[htbp]
\vspace{-2.5mm}
	\begin{minipage}{.5\linewidth}
		\centering
		\subfloat[]{\label{QT:a}\includegraphics[trim=5.5 1 1 1,clip,width=4.8cm]{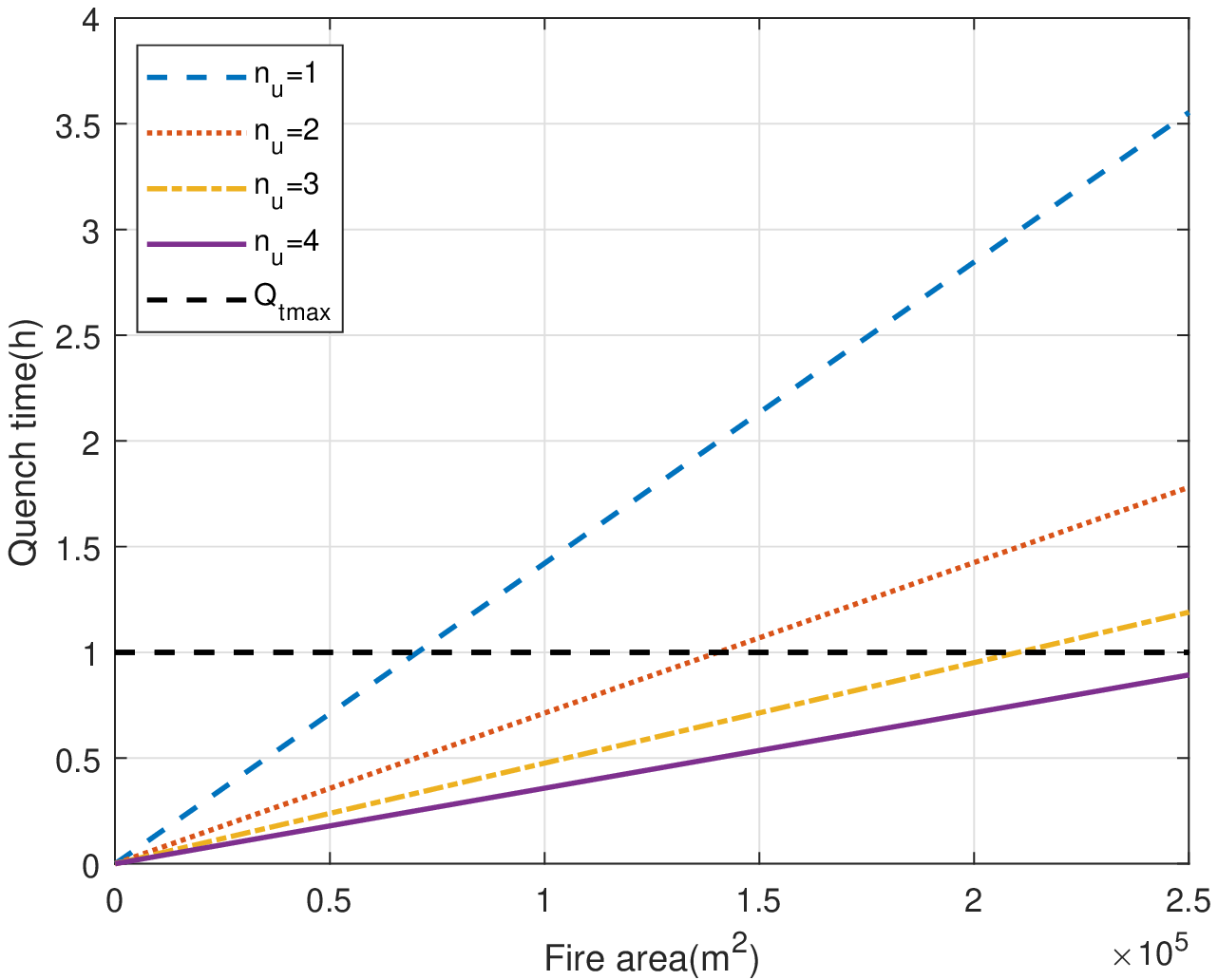}}
	\end{minipage}%
	\begin{minipage}{.5\linewidth}
		\centering
		\subfloat[]{\label{QT:b}\includegraphics[trim=3.5 1 4 1,clip,width=4.8cm]{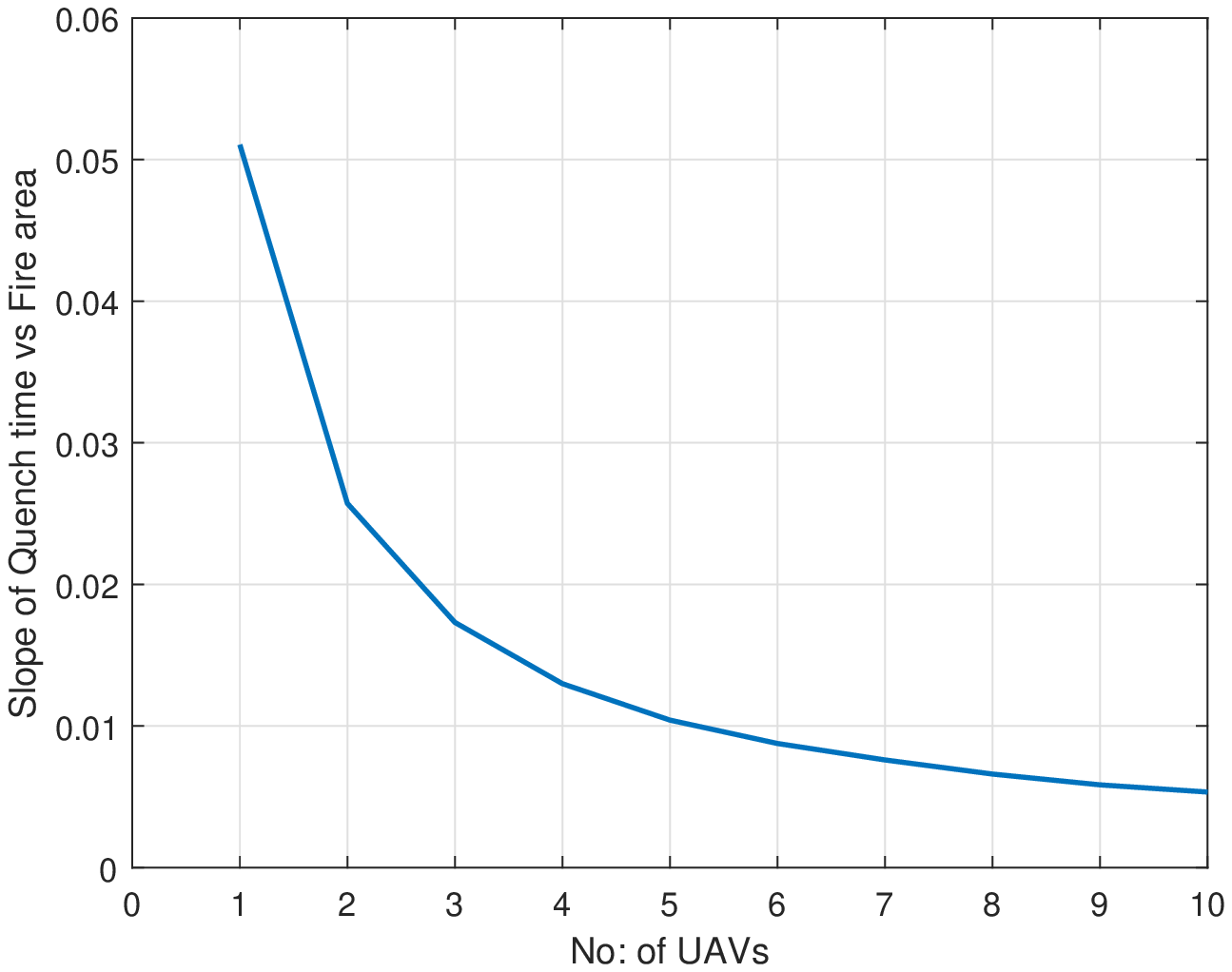}}
	\end{minipage}
	\caption{Plots showing (a) Variation of quench time with fire area for different number of UAVs. (b)Variation of slope for different number of UAVs.}
	\label{quenchvsarea}	
\end{figure}
The quench time is defined as the time taken to quench the fire completely after the detection of fire by any number of UAVs. The quench time $Q_t^j$ for the $j^\text{th}$ fire depends on the quench area rate ($r_q$), fire area to be extinguished ($A_f^j$), number of UAVs detecting the fire ($N_{qu}^j$), and the time of joining of UAVs acting on the fire ($T_m^j$).
\begin{equation}
	Q_t^j=f(r_q,A_f^j,N_{qu}^j,T_m^j)
\end{equation}
Let the UAVs spray water at a flow rate $W_r$ in \si{kg/s}, then the quench area rate ($r_q$) in (\si{m^{2}/s}) is given by
\begin{gather}
	r_q=\frac{W_r}{CF}
\end{gather}
The quenched area $A_q^j(t)$ at time $t$ \si{s} from the start of mission is given by
\begin{gather}	
	A_q^j(t)=\sum_{m=1}^{N_{qu}^j} r_q u(t-T_m^j)
	\end{gather}
The fire area $A_f^j$ (\si{m^{2}}) could be quenched in $Q_t^j$ (\si{s}) assuming same time of joining for $N_{qu}^j$ UAVs.
\begin{gather}
	Q_t^j=\frac{A_f^j}{N_{qu}^jr_q} 
\end{gather}	

Fig. \ref{QT:a} shows the variation of quench time with fire area for different numbers of UAVs with the same joining time. If multiple UAVs act on the same target, then the reduction in fire area multiplies with the number of UAVs, considerably reducing the quench time. The slope variation of the $Q_t$ vs $A_f$ graph with different number of UAVs is shown in Fig. \ref{QT:b}. The slope reduces with an increase in UAVs, but the reduction is minimal after a point. This explains the importance of multi-swarm missions with regulated merging and repulsion of swarms for the dynamic target scenarios.
\subsection{Performance Model}
The goal behind forest firefighting is to reduce the mission time, minimizing the destruction of biodiversity. The total time required to mitigate all the fire locations in the search area is defined as the mission time. This includes the detection time and quench time for all fire locations. The search objective of the mission is to maximize the number of targets detected in a given time, or equivalently, minimize the detection time. The mitigation objective is to minimize the quench time of detected targets. The active fire area increases with the increase in detection time. The minimization of detection time results in the smallest fire area to be quenched by the UAVs. Thus, the search objective can alternatively be represented as the minimization of area of the detected fire locations. Increasing the number of swarms acting on the same fire reduces quench time and achieves the mitigation objective, but the effect diminishes after a certain point. Furthermore, more swarms quenching the same fire reduce the detectability of undetected fires, increasing the detection time of undetected fire locations. The fire will spread, resulting in crown fires and a longer quench time. It is vital to balance the number of searching and quenching swarms for a successful mission. The search objective is split into minimization of the area of detected and undetected targets to achieve this goal. The multi-objective performance model can be represented using a weighted sum model as
\begin{gather}
	\nonumber
	\text{min}~ w_1\sum_{j=1}^{n_{df}} A_f^j+w_2\sum_{j=n_{df}+1}^{n_f} A_f^j+w_3 \sum_{j=1}^{n_{df}} Q_t^j\\
	\text{s.t} Q_t^j<Q_{tmax}
	\label{perfobj}
\end{gather}

where $n_{df}$ is the number of fire locations detected, $w_1$, $w_2$ and $w_3$ are the weights of the objective functions. The choice of weights determines the condition for repulsion and merging of swarms and helps maintain the balance between searching and quenching swarms. The MSCIDC algorithm is formulated to achieve the search and mitigation objectives of multi-swarm forest firefighting.
\section{Multi-Swarm Cooperative Information-driven search and Divide and Conquer mitigation control}
\begin{figure}[htbp]
	\centering		
	{\includegraphics[ width=8.7cm]{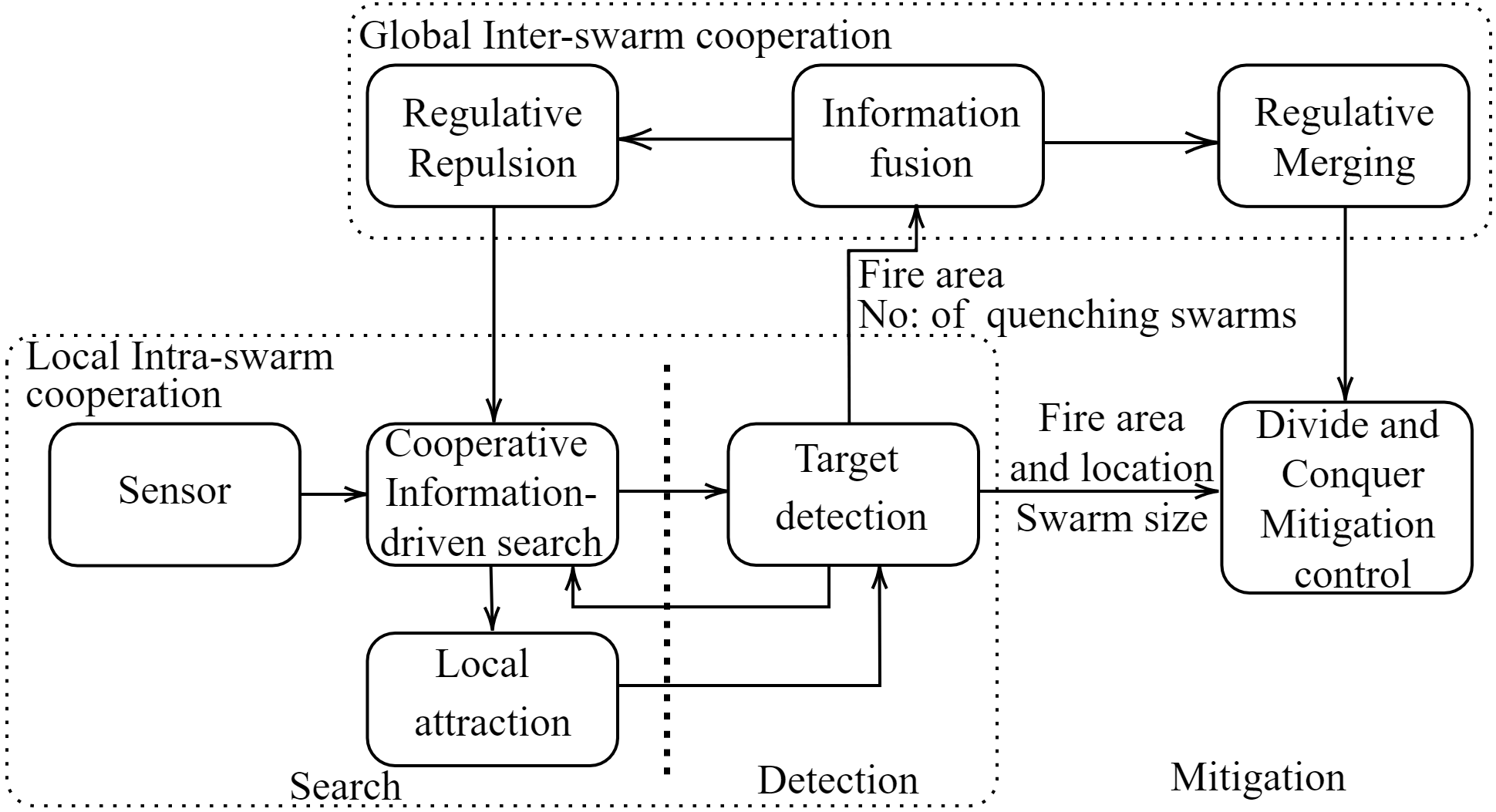}}%
	\caption{Schematic diagram of the proposed MSCIDC}
	\label{blockdia}
\end{figure}

The Multi-Swarm Cooperative Information-driven search and Divide and Conquer mitigation control (MSCIDC) uses swarms of UAVs to search and mitigate forest fire. The schematic diagram of the proposed method is shown in Fig. \ref{blockdia}. In the proposed method, swarms use multi-swarm cooperative information-driven search during the search phase of the mission. The search method combines cooperative information-driven exploration and exploitation for the faster detection of targets.
The cooperation between the members of the same swarm based on sensor information is termed local intra-swarm cooperation, which helps the members of the swarm search in the direction of maximum information and helps detect the same fire location by the non-detector members of the swarm. After detecting the fire location, swarm members fly in divide and conquer mitigation control to mitigate the fire by spraying water. The cooperation between swarms is global inter-swarm cooperation, which helps in the merging and repulsion between swarms during the mitigation of detected fires. The global inter-swarm cooperation leads to faster mitigation of detected fires and faster detection of other undetected fire locations.
\subsection{ Multi-Swarm Cooperative Information-driven Search}
Multi-swarm cooperative information-driven search is a multi-stage search process depending upon the temperature value sensed by the temperature detector. The two-stages of MSCIDC are cooperative information-driven exploration and exploitation. The balance between exploration and exploitation is a crucial factor in any stochastic search. The temperature threshold value helps in establishing the balance in the proposed method. The search process will switch from cooperative information-driven exploration to exploitation if the temperature sensed is higher than a threshold value.

The target search missions require initial identification and subsequent detection of targets. If the atmospheric temperature reading from the temperature detector exceeds a threshold, then a potential fire area is identified. Cooperative information-driven exploration is performed until the identification of a possible fire location. UAV swarm moves to the next stage of the search process once a potential target is identified. The swarm uses a cooperative information-driven exploitation in this stage and switches back to the cooperative information-driven exploration if the temperature falls below the threshold value.

\begin{figure}[htbp]
    \vspace{-3.5mm}
	\centerline{\includegraphics[width=7.8cm]{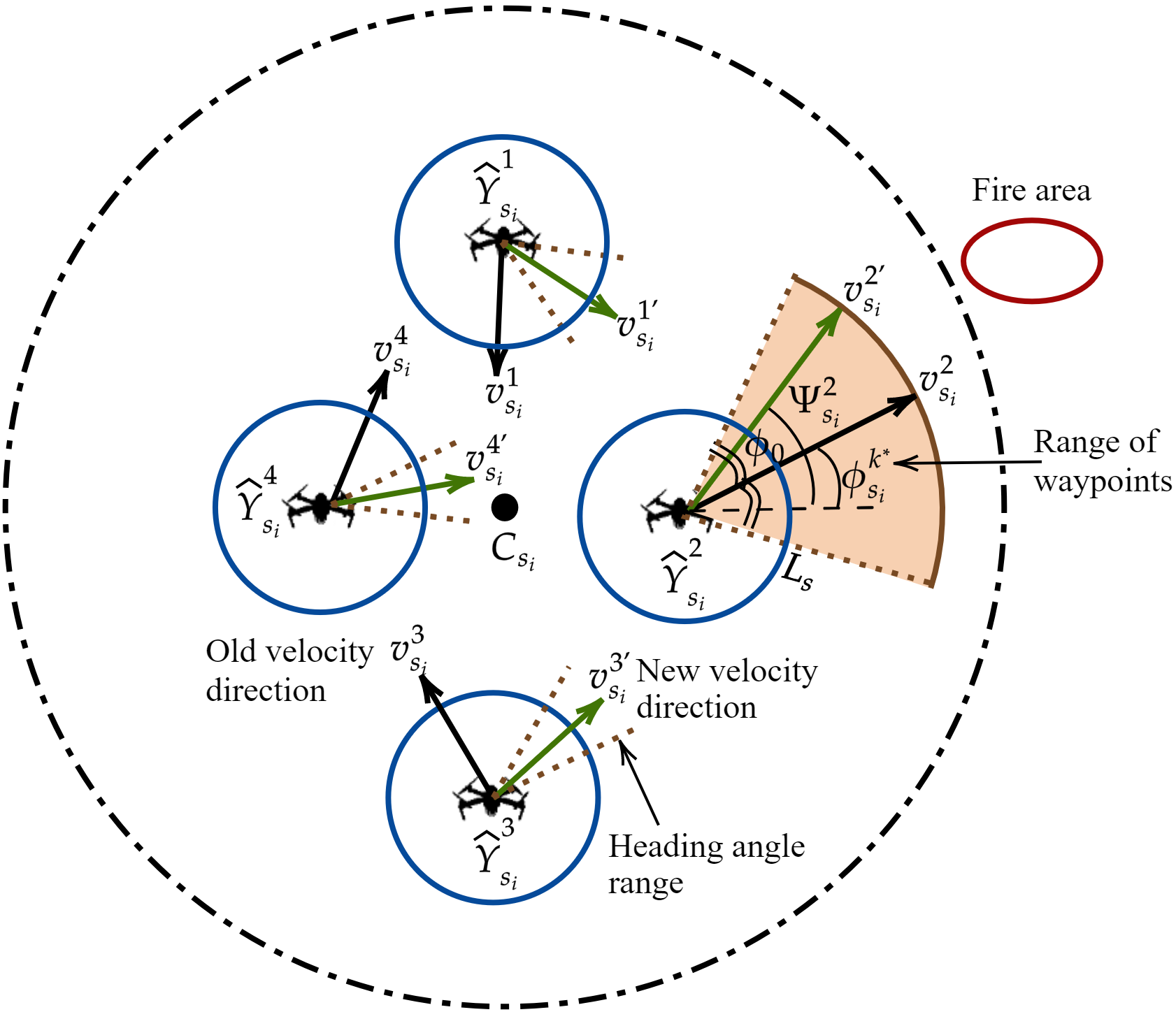}}
	\caption{Working of cooperative information-driven search.}
	\label{IDLschematic}
    \vspace{-1.5mm}
\end{figure}

The working of cooperative information-driven search is shown in Fig. \ref{IDLschematic}. The swarms are desired to move in the direction of the maximum temperature gradient sensed by UAVs in a swarm. The movement of swarms to the desired location depends on the cooperative action of all swarm members. The cooperative behavior of swarm members attributes to the shift of the swarm center. Communication between swarm members is essential to maintain the swarm structure and share the information sensed about the fire location to ensure the cooperative mission.
%According to the search algorithm, the new waypoint generated will lie in the shaded region.

 The swarm members broadcast their positions, temperature, rate of temperature, the probability of detection, and the fire location specifications sensed by each member within the swarm. Thus $k^\text{th}$ swarm member of the $s_i^\text{th}$ swarm will have data $ \hat{Y}_{s_i}^{\,k}(t)=(p_{s_i}^{\,k},T_{s_i}^{\,k},\frac{dT_{s_i}^{\,k}}{dt},P_j^{\,k})$ from the sensors for the corresponding position of UAV. If the $P_j^{\,k}>\gamma$, then $j^\text{th}$ fire is detected and the UAV have the information $\hat{Y}_{s_i}^{\,k}(t)=(p_{s_i}^{\,k},T_{s_i}^{\,k},\frac{dT_{s_i}^{\,k}}{dt},P_j^{\,k},C_{f_j},a_j,b_j,h_j^{\,k})$ where $h_j^{\,k}$ is the heading direction towards the $j^\text{th}$ fire from the $k^\text{th}$ UAV. Each swarm member broadcasts the sensed information to all the members in the swarm. The information structure available for the $s_i^\text{th}$ swarm is $\hat{Y}_{s_i}(t)=\hat{Y}_{s_i}^1\cup \hat{Y}_{s_i}^2....\cup \hat{Y}_{s_i}^{n_{s_i}}$. The information structure is used by the swarm to push the swarm center towards the desired direction. The swarm member with maximum information, $k^\ast$ is defined as the member having the highest temperature gradient.
\begin{equation}
	k^\ast=\argmax\limits_{(k\in s_i)}\frac{d}{dt}(T_{s_i}^{\,k})
\end{equation}
The maximum information direction, $\phi^{k^\ast}_{s_i}$ is defined as the velocity direction of swarm member having maximum information and $\hat{Y}^{k^\ast}(t)$ is used to drive the swarm in maximum information direction. In Fig. \ref{IDLschematic}, UAV $2$ has a velocity direction towards the fire and has the highest temperature gradient, hence $k^\ast=2$. The range of waypoints generated for swarm members in the search area is shown as a shaded region. The heading angle range of swarm members depends on the range of waypoints. Therefore, the search space of swarm member is constrained to heading angle range. The search angle $\Psi^{\,k}_{s_i}$ for the $k^\text{th}$ swarm member to generate a waypoint in the maximum information direction is drawn from a uniform distribution from $\phi^{k^\ast}_{s_i}-\phi_0$ to $\phi^{k^\ast}_{s_i}+\phi_0$.
\begin{gather}
	\Psi^{\,k}_{s_i}\sim U({\phi^{k^\ast}_{s_i}-\phi_0, \phi^{k^\ast}_{s_i}+\phi_0})
 \end{gather}
The angle $\phi_0$ is calculated using 
\begin{gather}
	\phi_0=\frac{K_\phi}{1+exp(-K_eT_s)}
	%T_s= \max\limits_{(k\in s_i)}(T_{s_i}^{\,k})%
\end{gather}
where $T_s$ is the maximum temperature sensed by the swarm, $K_\phi$ and $K_e$ are constants. The new waypoint for $k^\text{th}$ swarm member, $p_{rs_i}^{\,k}(t)$ is calculated as
\begin{gather}
	p_{rs_i}^{\,k}(t)=p_{s_i}^{k^\ast}(t)+L_s*l^{\,k}_{s_i}	
	\begin{bmatrix}
		cos(\Psi^{\,k}_{s_i})&sin(\Psi^{\,k}_{s_i})
	\end{bmatrix}
	\label{stochsearch}
\end{gather}
where $p_{s_i}^{k^\ast}(t)$ is the position of swarm member with maximum information, and the second term in Eq.\ref{stochsearch} corresponds to the movement contributed by the stochastic search process. $L_s$ is the search step length, $l^{\,k}_{s_i}$ is individual lengths drawn from the probability distribution functions. The waypoints can be generated using different probability distribution functions. The appropriate selection of these functions results in effective exploration and exploitation of the search area. The swarm members are navigated towards the generated point, and only after reaching the current waypoint, a new waypoint is generated. 
%The heading angle range of swarm member depends on the range of waypoints. Therefore, the search space of swarm member is constrained to heading angle range for each swarm member.
\subsubsection{Cooperative Information-driven Exploration}
The cooperative information-driven exploration is performed as long as the maximum temperature sensed by the swarm, $T_s$ is less than the threshold value. 
\begin{gather}
	T_s<\xi
\end{gather}
where $\xi$ is the detection threshold of temperature. In the case of information-driven exploration, levy distribution is used to calculate the stochastic search factor. The search step length $L_s$ in Eq.\ref{stochsearch}
% \begin{equation}
% 	L_s=s_l
% \end{equation}
 is the levy step length, and $l^{\,k}_{s_i}$ is the individual lengths of the levy flight legs, which are drawn from the Levy distribution. The cooperative information-driven exploration uses higher step lengths, which helps in the exploration of the search area. When $\phi_0=\pi$, the search becomes a conventional levy search. The heading angle and randomized leg lengths compose to form the cooperative information-driven exploration trajectory.
\subsubsection{Cooperative Information-driven Exploitation}
The cooperative information-driven exploitation is used for exploiting the search area once the UAV swarm reaches a possible target location. The possible target location means a location with a temperature greater than the detection threshold.
% \begin{equation}
% 	T_s\ge \xi
% \end{equation}	
For cooperative information-driven exploitation, Brownian step length is used for better exploitation of the region, and $l^{\,k}_{s_i}$ in Eq.\ref{stochsearch} is drawn from normal distribution $N(0,1)$.
% \begin{gather}
% 	\nonumber
% 	L_s=s_b
% \end{gather}
 The step length used in this exploitative search is considerably shorter compared to the explorative search, $s_b<<<s_l$. The shorter step length aids in the efficient exploitation of the search area to reach the precise target location.
\subsubsection{Local attraction}
The attractive force acting within the swarm is termed as Local attraction, and it is the reason for the preservation of swarm structure in the search and detect phase. The local attraction comes into play either when a swarm member moves outside the swarm boundary or when a swarm member detects a fire. If any UAV in a swarm tends to move outside the swarm boundary, the attractive force of the swarm center pushes the UAV inside. A fire location is detected if the probability of detection determined using Eq. (\ref{probconf}) exceeds the detection threshold ($\gamma$). The swarm member with maximum information detects the fire and attracts other swarm members towards it. The attraction towards the fire-detected member helps the other swarm members to reach the fire front faster. After target detection, Divide and conquer mitigation control is employed for target mitigation.

\subsection{Divide and Conquer Mitigation Control}
Dynamic targets like forest fires require uninterrupted attention to stop the spread and for accelerated mitigation. The swarms should start quenching the fire along the fire front to prevent spreading after detecting the fire location. The mitigation control is designed to cover an equal non-overlapping area by UAVs detecting fire. The members of the swarm should cover an equal non-overlapping area even when the area is shrinking to avoid the buffer time regarding the overlapped area.
\begin{figure}[htbp]
\vspace{-2.5mm}
	\centerline{\includegraphics[width=60mm]{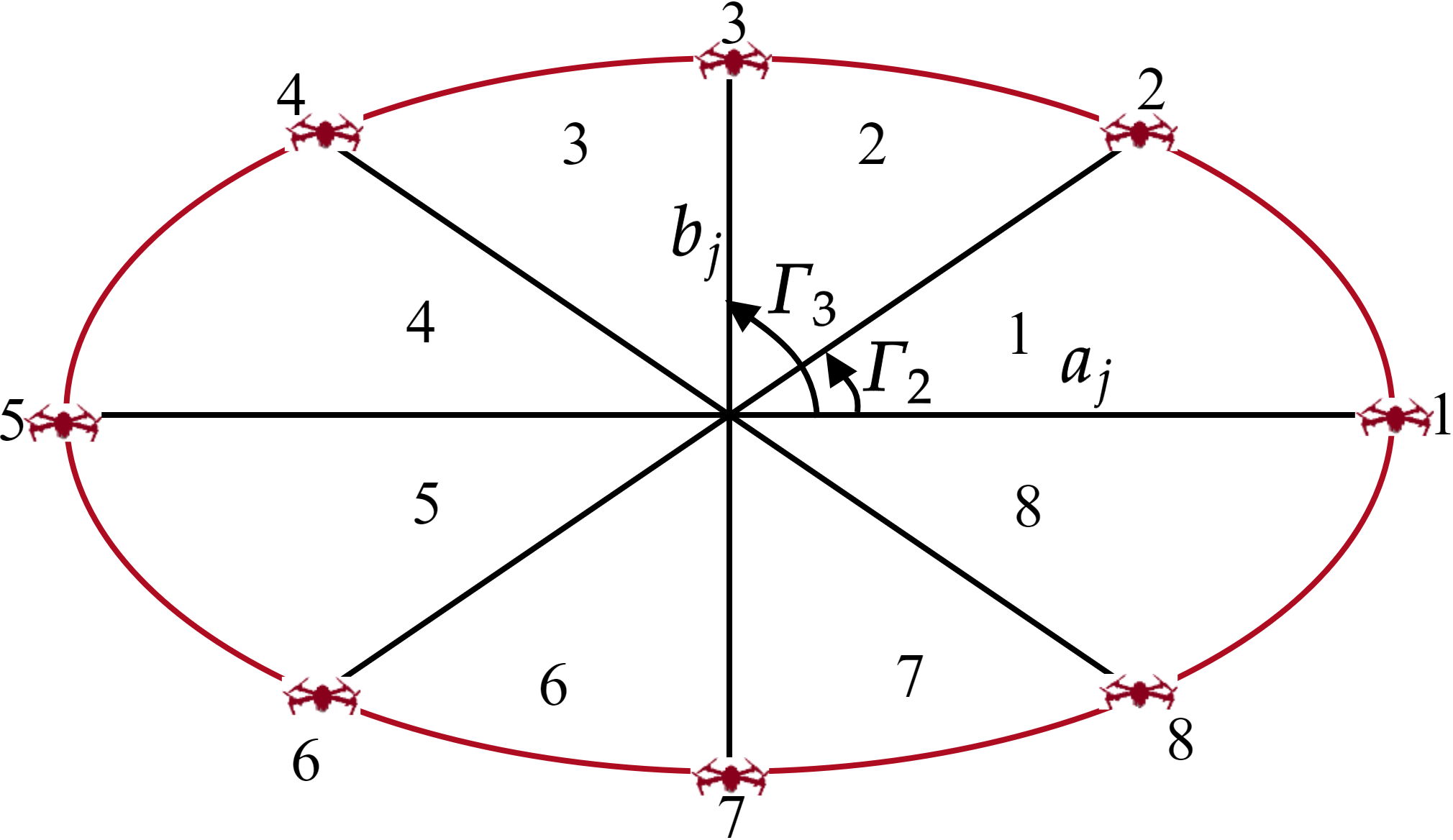}}
	\caption{Divide and conquer mitigation sectors for 8 UAVs for an elliptical fire front.}
	\label{formation}

\end{figure}
Each UAV in the swarm is restricted to quench an elliptical or circular sector of the equal fire area. Fig. \ref{formation} shows the initial alignment of 8 UAVs according to the proposed mitigation control. The UAVs move in to and fro motion in the allotted sector along the firefront.	

Let $N_{qu}^j$ be the total number of UAVs in the $N_{qs}^j$ swarms quenching the $j^\text{th}$ fire. The angular position and reference angular position of the $m^\text{th}$ UAV at time $t$ be $\theta_{s_i}^{\,m}(t)$ and $\theta_{rs_i}^{\,m}(t)$ where $m = 1, 2,..., N_{qu}^j$. The total fire area is divided into $N_{qu}^j$ equal area sectors. The area of $m^\text{th}$ sector is
\begin{multline}
	A_{m*m+1}(t)=\frac{a_jb_j}{2}\bigg[\taninv\bigg(\frac{a_j\tan\Gamma_{m+1}(t)}{b_j}\bigg)\\-\taninv\bigg(\frac{a_j\tan\Gamma_{m}(t)}{b_j}\bigg)\bigg]
	\label{nonoverlappingarea}
\end{multline}
The sector angle, $\Gamma_m$ will be same for all sectors in the case of circular profile but different for the elliptical profile. The sector angle of the first sector, $\Gamma_1=0$, and $\Gamma_m$ for $m\ge2$ is calculated to get equal-area sectors. The control law for the $m^\text{th}$ UAV is given by
\begin{gather}
	\nonumber
	\dot{\theta_{s_i}^{\,m}}(t)=\omega(t)+K_m(\theta_{s_i}^{\,m}(t)-\theta_{rs_i}^{\,m}(t))\\
	\dot{\theta_{rs_i}^{\,m}}(t)=\mu(t)\omega(t)
\end{gather}
where $\omega(t)$ is the nominal angular velocity of the UAVs, and the control law asymptotically tracks the reference angular displacement for gain values, $K_m<0$. The factor $\mu(t)$ controls the direction of motion of the UAV along the firefront.
\begin{gather}
	\mu(t)=
	\begin{cases}\label{F2}
		1& \text{if $\theta_{rs_i}^{\,m}(t)-\Gamma_{m-1} <\delta_\theta$ and}\\\quad& \text{sgn$(\dot{\theta_{rs_i}^{\,m}}(t))=-1$ }\\
		-1& \text{if $\Gamma_{m}-\theta_{rs_i}^{\,m}(t)<\delta_\theta$ and}\\\quad& \text{sgn$(\dot{\theta_{rs_i}^{\,m}}(t))=1$}
	\end{cases}
\end{gather}
The waypoint for the $m^\text{th}$ UAV mitigating $j^\text{th}$ fire location is generated as follows
\begin{gather}
	p_{rs_i}^{\,m}(t)=C_{f_j}+\begin{bmatrix}
		a_j(t)\cos\theta_{s_i}^{\,m}(t)&b_j(t)\sin\theta_{s_i}^{\,m}(t)
	\end{bmatrix}^T
\end{gather}
\begin{algorithm}[t]
	\caption{MSCIDC Algorithm}\label{alg:one}
		Initialize with $n_s$ swarms with $n_{s_i}$ swarm members\;
		Initialize $N_{qu}^j=0,N_{qs}^j=0$ for $j=1\text{ to }n_{f}$\;
		\For{$i=1\text{ to }n_s$}
		{Each $i^\text{th}$ swarm has information $\hat{Y}_{s_i}(t)$\;
		\For{$j=1\text{ to }n_{f}$}{
		\For{$k=1\text{ to }n_{s_i}$}{		
		\eIf{$P_j^{\,k}<\gamma$}
		{\eIf{$T_s<\xi$}
		{Cooperative information-driven exploration\;}
		{Cooperative information-driven exploitation\;}}
		{$k^\text{th}$ swarm member of $s_i^\text{th}$ swarm detects $j^\text{th}$ fire\;
% 		\If{$k\in s_i$}
% 		{$N_{qs}^j$=1
% 		}
		$N_{qu}^j=N_{qu}^j+1$\;
		Local Attraction of swarm members\;
		\For{$m=1\text{ to }N_{qu}^j$}	
		{Move to the initial alignment position of $N_{qu}^j$ member\; 
		Quenches $N_{qu}^j$ sector\;
		\eIf{$(A_f^j>\delta_A \text{ or }F_r<\delta_f)\text{ and }N_{qs}^j<\delta_s$}
		{$N_{qs}^j=N_{qs}^j+1$\;
		Global regulative merging}
		{\If{$k\notin s_i \text{ and }\gamma_0<P_j^{\,k}< \gamma$}{Global regulative repulsion}}
		}
		}
		}	
		}
		}
\end{algorithm}
\subsubsection{Global Regulative Merging and Repulsion}
The regulative merging of swarms helps in reducing the quench time, when multiple swarms detect the same fire location. The regulative merging of swarms happens either if the fire area detected is higher than a threshold, $\delta_A$ or if the number of remaining active fires, $F_r$, is substantially lesser. The total number of swarms allowed to merge, $N_{qs}^j$ is limited to $\delta_s$ to minimize the detection time of undetected targets. The condition for regulative merging can be represented as
\begin{equation}	
	(A_f^j>\delta_A \text{ or }F_r<\delta_f)\text{ and }N_{qs}^j<\delta_s
	\label{mergecond}		
\end{equation}
The threshold values of area $(\delta_A)$, remaining fires $(\delta_f)$ and number of swarms under mitigation $(\delta_s)$ are chosen to achieve the performance objectives of forest firefighting in Eq.\ref{perfobj}. This minimizes the detection time and quench time of detected fire based on the total number of swarms and its capacity. It is assumed that the approximate number of fire locations is known towards the end of the mission due to the effective coverage of the search area. The merged swarms split after completely quenching the fire and split swarms search for other potential fire areas.

The introduction of regulative repulsion of swarms reduces the mission time for the scenarios that don't satisfy the condition in Eq. \ref{mergecond}. If a swarm detects a fire under mitigation, the second swarm gets repelled by the first swarm. The repulsion happens when the probability of detection value of the second swarm is $\gamma_0<P_j^{\,k}< \gamma$. The repelled swarm uses cooperative information-driven exploration in the direction opposite to the maximum information direction. This repulsive action between swarms helps identify other unattended fire locations at an earlier stage. The dynamic swarm approach helps to maximize the target detection minimizing the detection and quench time. The algorithm of the proposed method is summarized in Algorithm \ref{alg:one}.

	\begin{figure*}[t]
	\vspace{-5.5mm}
		\centering
		\subfloat[ Simulation frame at time $1 \si{s}$] {\label{sim:a}
		\includegraphics[trim=8 5 0 5,clip,width=4.45cm]{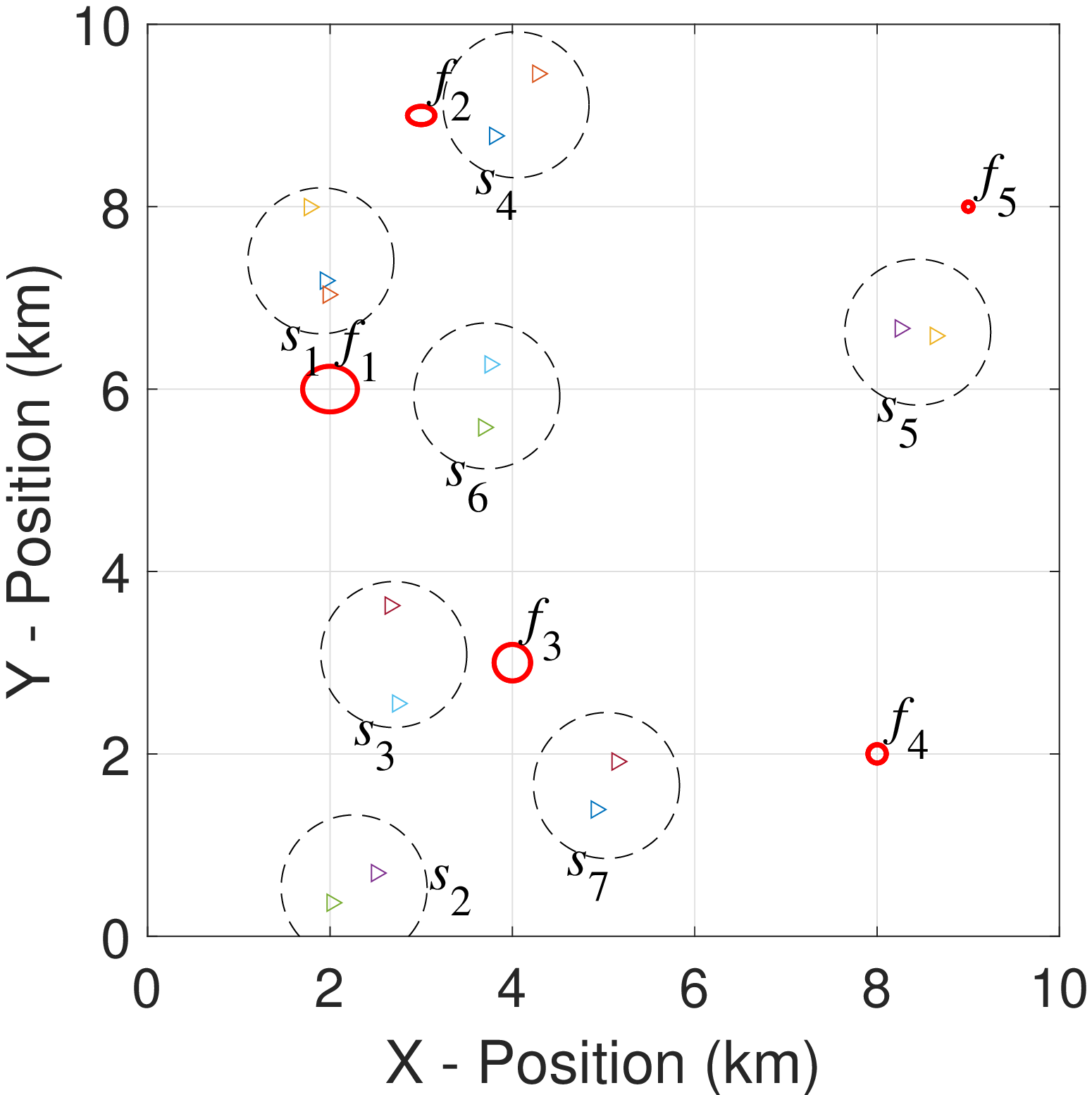}}%
		\subfloat[Simulation frame at time $112 \si{s}$]{\label{sim:b}
			\includegraphics[trim=8 5 0 5,clip,width=4.45cm]{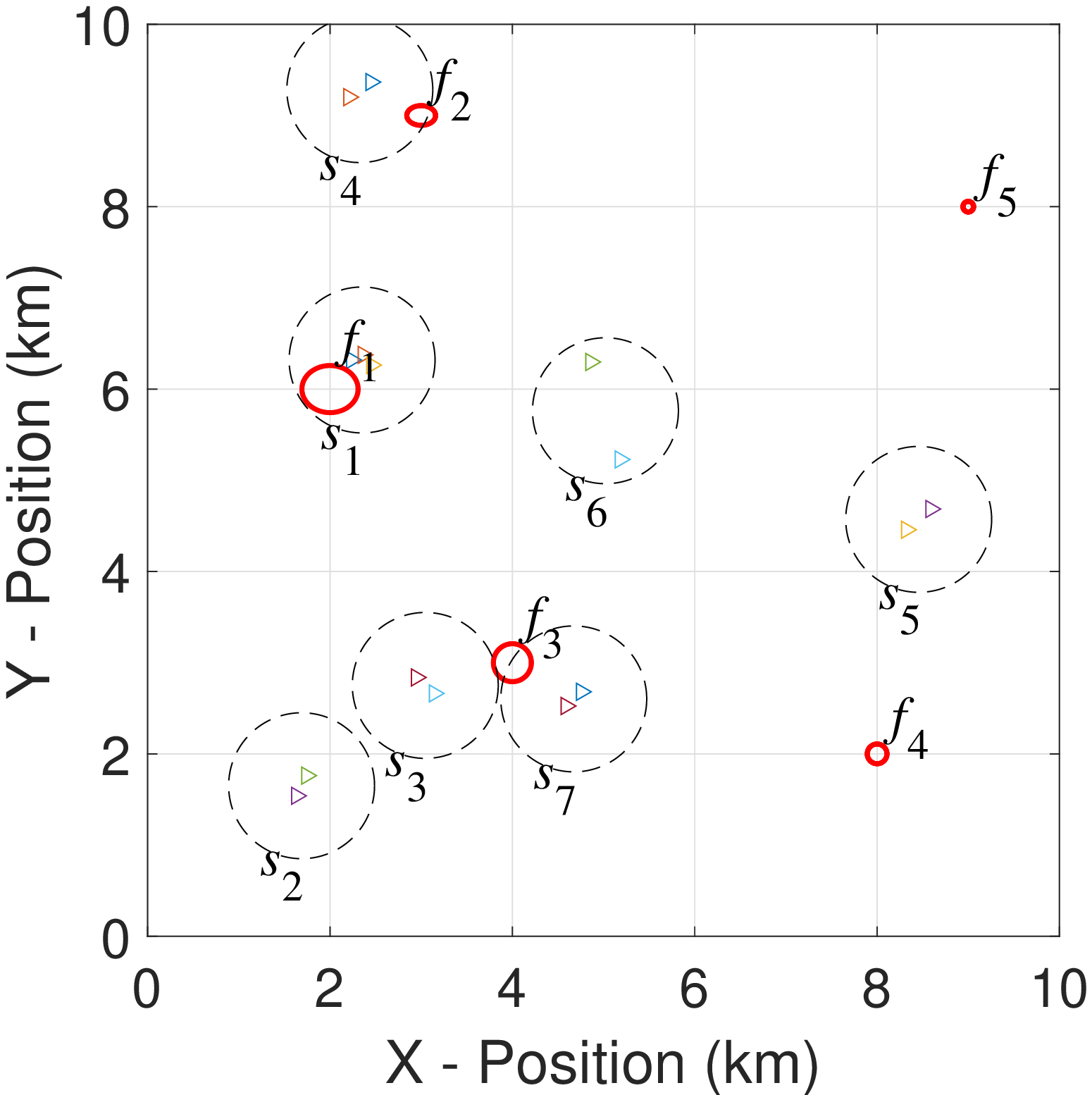}}%
		\subfloat[Simulation frame at time $420 \si{s}$]{\label{sim:c}
			\includegraphics[trim=8 5 0 5,clip,width=4.45cm]{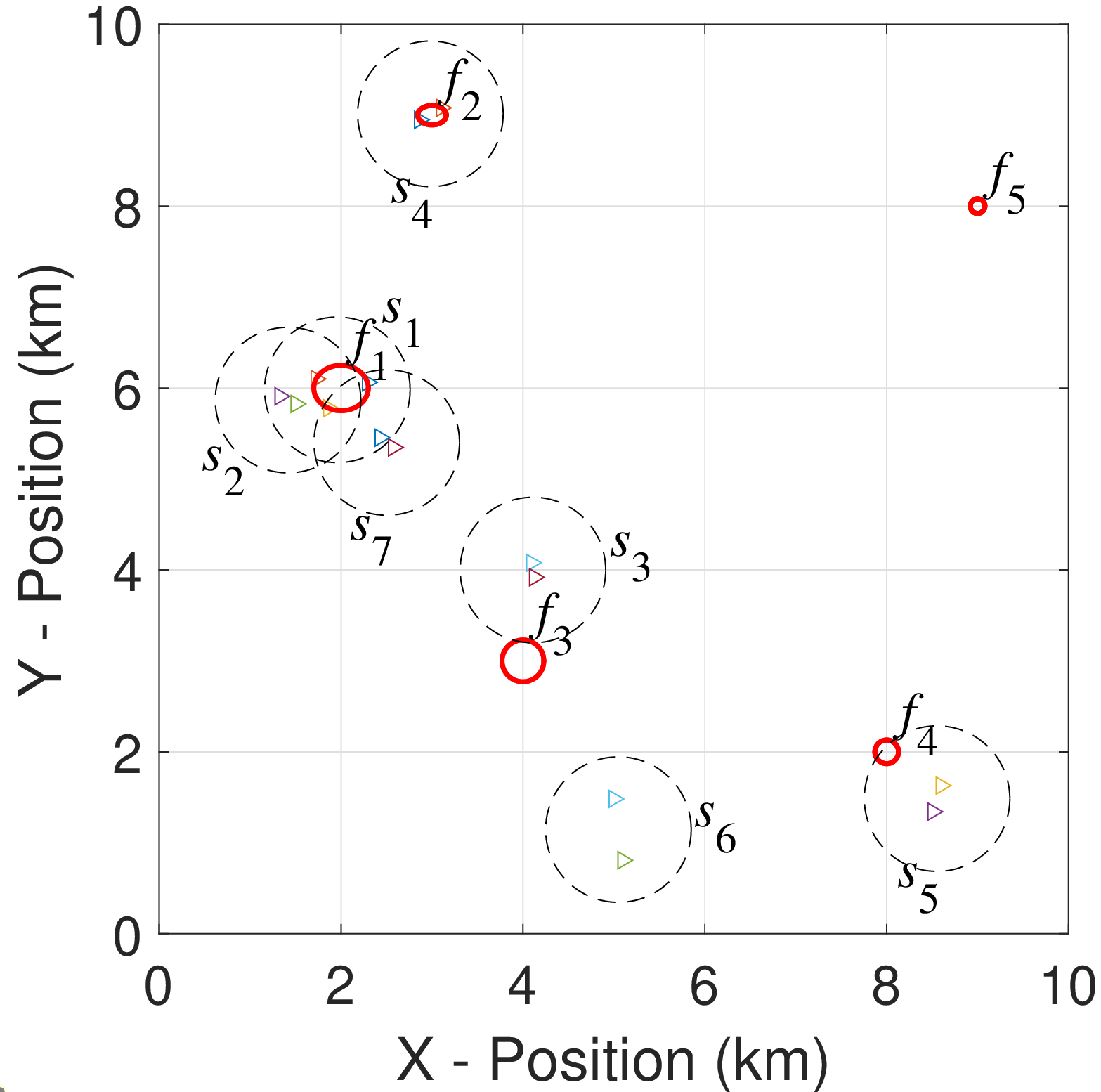}}%
		\subfloat[Simulation frame at time $1581 \si{s}$]{\label{sim:d}
			\includegraphics[trim=8 5 0 5,clip,width=4.45cm]{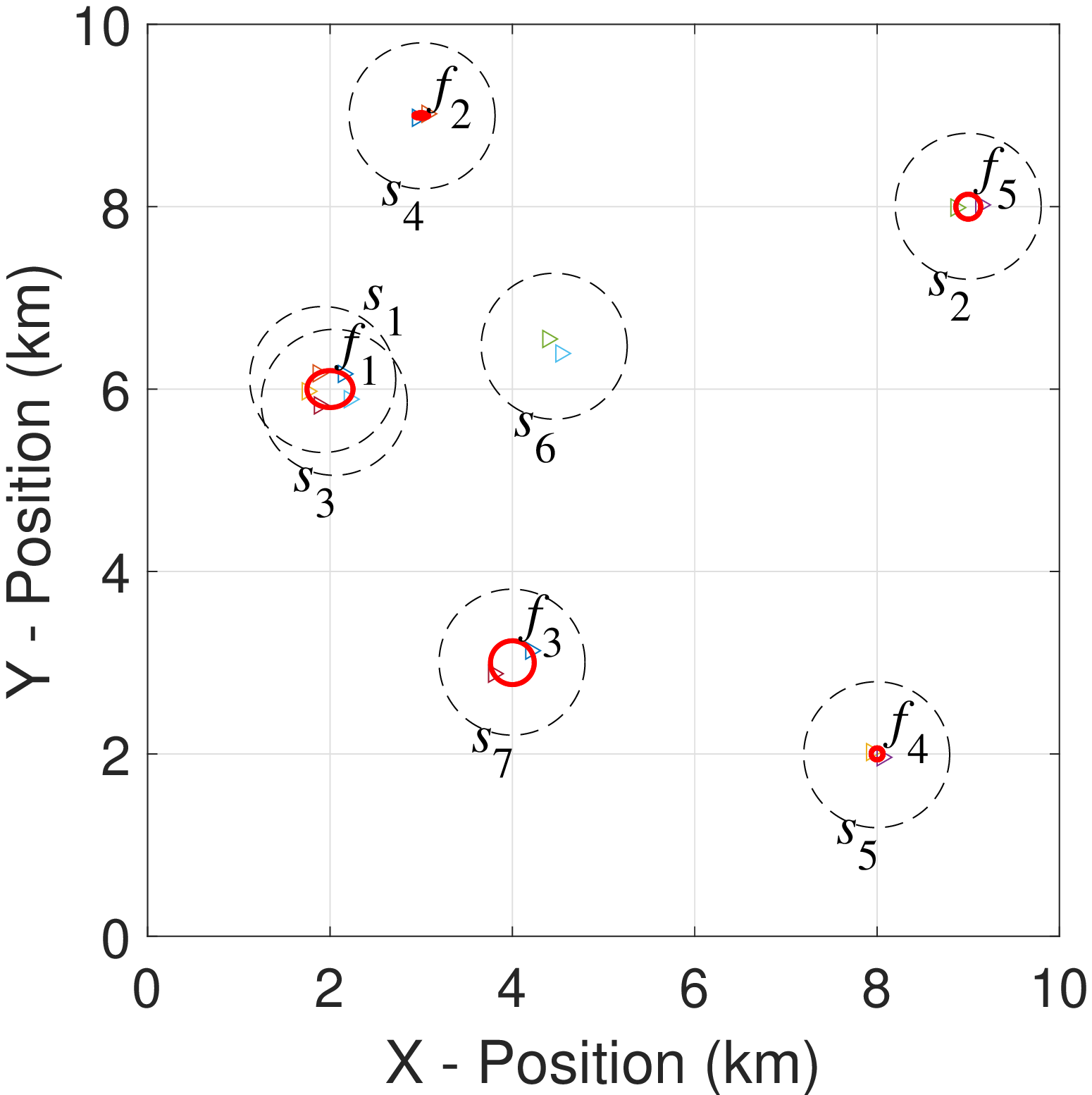}}%
		\caption{Working of MSCIDC: (a) All swarms performing cooperative information-driven search task (b) Swarm, $s_1$ detects fire, $f_1$ and UAVs in $s_1$ have local attraction, (c) $s_1$ starts quenching $f_1$ using Divide and Conquer mitigation control and swarms $s_2$ and $s_7$ get repelled due to global regulative repulsion (d) Global regulative merging of $s_1$ and $s_3$ to mitigate $f_1$.}
		\label{mscidcsteps}
		\vspace{-2.5mm}
	\end{figure*}

\section{Numerical Simulation Results}
The proposed MSCIDC is analyzed for an area of size $10$ \si{km} x $10$ \si{km} considering 15 UAVs with varying swarm sizes for detection and mitigation of five fire locations. Two of the five fire locations have a circular profile, and the rest have an elliptical profile. The fire coordinates and initial lengths of semi-major and semi-minor axes are given in Table \ref{fireloc}. The members of swarm are flying with a constant speed of $20$ \si{m/s} during the search. During mitigation, the speed is varied according to the size of the fire area. The swarms initiate the search from different locations within the search area under consideration.
	\begin{table}[htbp]
	\caption{Fire Locations and Initial Size}
	\centering
	\begin{tabular}{|c|c|c|c|}\hline
		Sl.No & $C_{f_j}$ (\si{m}) & $a_j$ (\si{m}) & $b_j$ (\si{m})\\ \hline
		1 & (2000, 6000) & 300 & 250\\ \hline
		2 & (3000, 9000) & 150 & 100\\ \hline
		3 & (4000, 3000) & 200 & 200\\ \hline
		4 & (8000, 2000) & 100 & 100\\ \hline
		5 & (9000, 8000) & 50 & 50\\ \hline
	\end{tabular}
	\label{fireloc}
	\end{table}
The simulations are performed in MATLAB R2021a environment with an Intel Core-i$7$, $3.2-$GHz processor and $16-$GB memory. The Monte-Carlo analysis is performed for $100$ iterations to evaluate the average performance of the MSCIDC by varying the number of swarms for the same number of UAVs. The results are compared with the Oxyrrhis Marina-inspired Search (OMS) \cite{OMS}, and other random search methods with different probability distributions to analyze the efficiency of the proposed method. The results of the proposed method are compared with different multi-UAV search methods with Dynamic Formation Control (DFC) used in \cite{OMS} for fire mitigation.

The performance indices used to analyze the results are detection time, mission time, and Fire Expansion Ratio (FER). Detection time is the time required to detect all the targets, whereas mission time is the time taken to complete the mission, i.e., the total time taken to detect and mitigate all the fires. The FER is an index to evaluate the extent of burnt area. It is the ratio of the increase in fire area after the commencement of the mission to the initial fire area.
%The increase in the fire area during the mission is calculated as a percentage of the initial fire area to evaluate the burnt forest area.

The working of the MSCIDC is explained for a scenario of 15 UAVs searching for the five fire locations given in Table \ref{fireloc}. The 15 UAVs are grouped into $7$ swarms with a swarm size of $\begin{bmatrix}
	3&2&2&2&2&2&2
\end{bmatrix}$. The different features in the proposed method can be analyzed from Fig. \ref{mscidcsteps}. The swarms are initiated randomly in the search area at the beginning of the mission. In Fig. \ref{sim:a}, the swarms are in the search phase of the mission to detect the fire location using cooperative information-driven search. When the swarms are far from the fire, the members use cooperative information-driven exploration to spot potential fire spots, and when the swarms get closer to the fire, they use cooperative information-driven exploitation. In Fig. \ref{sim:b}, swarm $s_1$ detects the fire location $1$ at $(2000, 6000)$ in the search area. The swarm member with a probability of detection greater than the detection threshold confirms the fire location and moves towards the waypoint above the fire front for mitigation. Subsequently, other swarm members are attracted towards fire location $1$ due to local attraction, leading to the detection of the same fire location. Fig. \ref{sim:c} shows the divide and conquer mitigation control of fire location 1 by the swarm members. Each member of the swarm mitigates the fire by spraying water in the assigned sector of the fire area. The swarms $s_2$ and $s_7$ moving closer to fire $1$ get repelled due to the global regulative repulsion in Fig. \ref{sim:c}. The swarms $s_2$ and $s_7$ repelled from fire $1$ detects fires $5$ and $3$ respectively. The global regulative repulsion helps in the faster detection of undetected fire locations. All the fire locations are detected due to search area coverage of the swarms towards the end of the mission, and the condition for global regulative merging is satisfied. In Fig. \ref{sim:d}, two swarms $s_1$ and $s_3$ merge to mitigate fire $1$ to reduce the quench time.

The proposed method is evaluated using Monte Carlo simulations for a different number of swarms with the same number of UAVs. The case with three swarms and five swarms have an equal number of UAVs in all swarms. The cases with $6$ and $7$ swarms have a swarm size of $\begin{bmatrix}
	3&3&3&2&2&2
\end{bmatrix} $ and $\begin{bmatrix}
	3&2&2&2&2&2&2
\end{bmatrix}$ respectively.

\begin{table}[h]
\caption{Performance of MSCIDC for different swarm numbers}
	\centering
 \begin{tabular}{|c|c|c|c|}
\hline
\begin{tabular}[c]{@{}c@{}}Number of\\ swarms\end{tabular} & \begin{tabular}[c]{@{}c@{}}Mean detection \\ time (min)\end{tabular} & \begin{tabular}[c]{@{}c@{}}Mean mission \\ time (min)\end{tabular} & Mean FER \\ \hline
3 & 41.82 & 107.01 & 1.029 \\\hline
5 & 31.03 & 81.19 & 0.642 \\\hline
6 & 27.59 & 83.27 & 0.550 \\\hline
7 & 24.13 & 71.56 & 0.477 \\ \hline
\end{tabular}
\label{varyingswarm}
\end{table}

 Table \ref{varyingswarm} gives the results for MSCIDC for different swarm numbers. The average detection time decreases with an increase in the number of swarms. More the number of swarms, more distributed the total UAVs as swarms. The regulative repulsion also helps in detecting unattended fire location reducing the mission time and improving the performance. The $80\%$ of total fire locations are detected in $15\%$ of the total mission duration for $7$ swarm case. Even though the average mission time decreases with an increase in number of swarms, the effect declines beyond a number. The FER also decreases with increase in number of swarms as fire locations are detected at an earlier stage. Thus the $7$ swarm case has the lowest mean detection time, mission time, and destroyed area.
 
\begin{figure}[htbp]
    \vspace{-5mm}
    \centering
	\subfloat[{} \label{sssq1}]
	{\includegraphics[trim=5 5 0 5,clip,width=4.5cm]{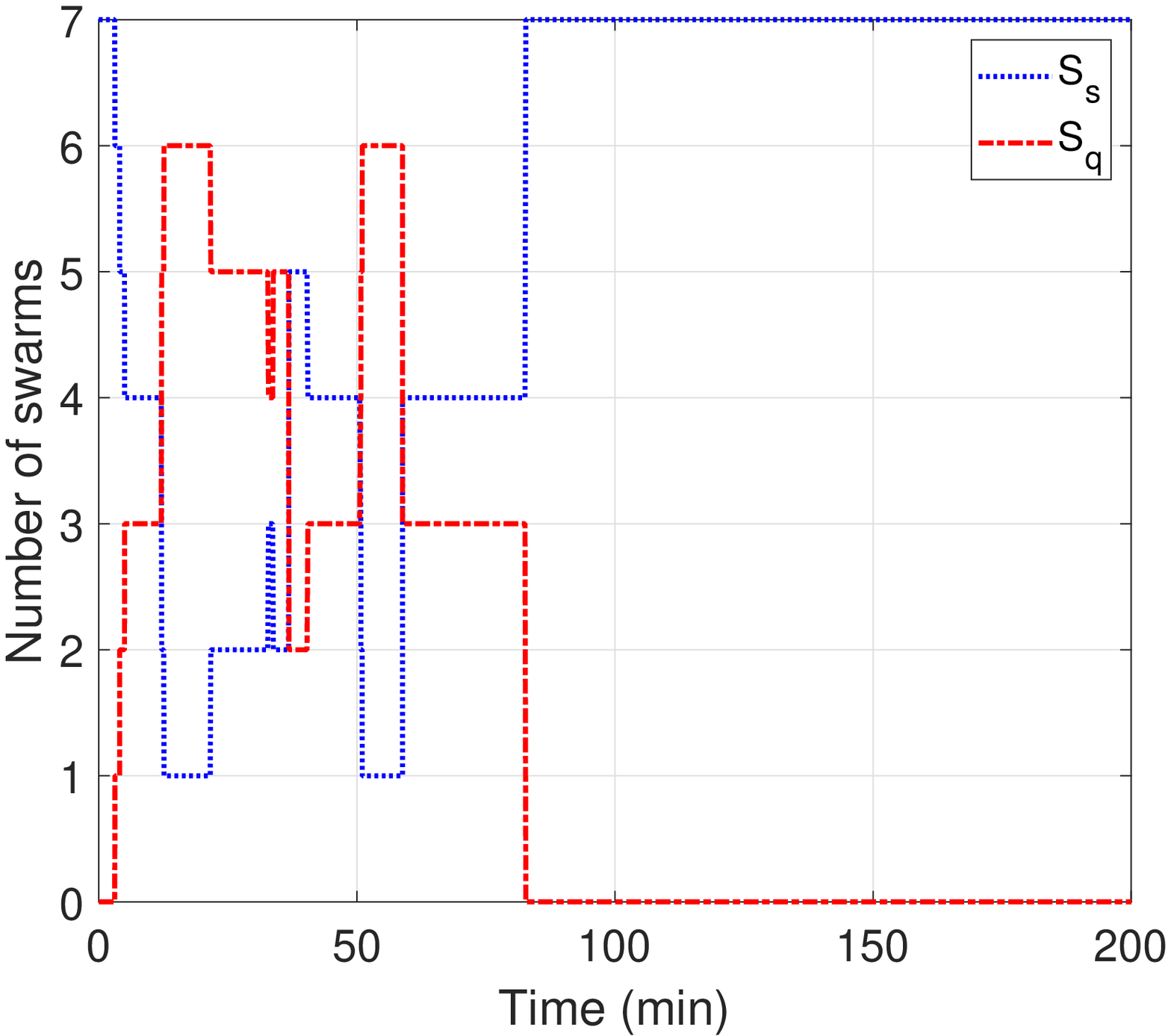}}%
		\subfloat[{} \label{sssq2}]
	{\includegraphics[trim=5 5 0 5,clip,width=4.5cm]{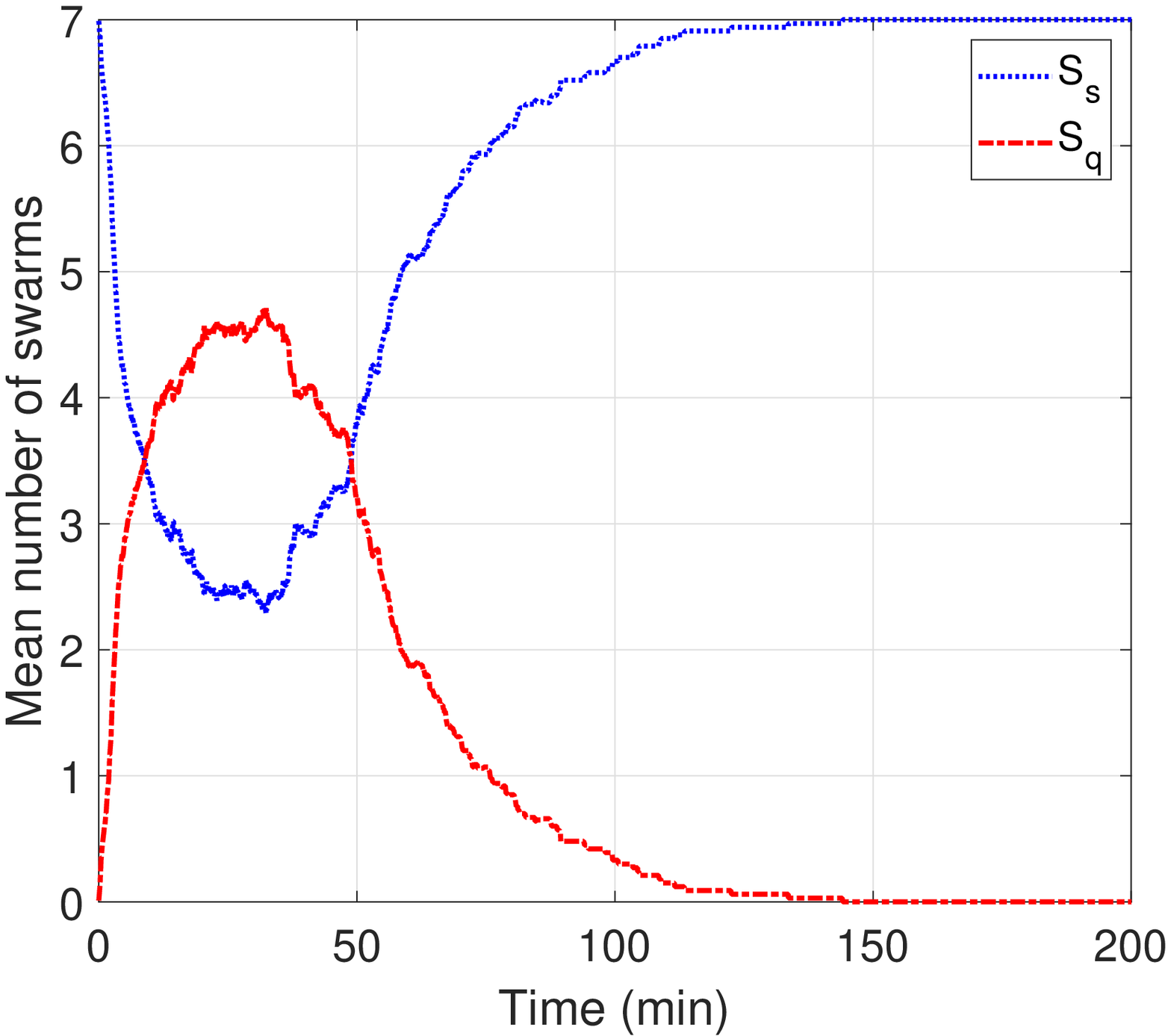}}%
	\caption{Number of swarms performing search ($S_s$) and mitigation ($S_q$) for (a) single run (b) Monte-Carlo simulation.}
	\label{SSSQ}
	 \vspace{-2.5mm}
\end{figure}
\begin{figure}[htbp]
    \vspace{-2.5mm}
	\centering
	\subfloat[{} \label{fffdfr1}]
	{\includegraphics[trim=5 5 0 3,clip,width=4.5cm]{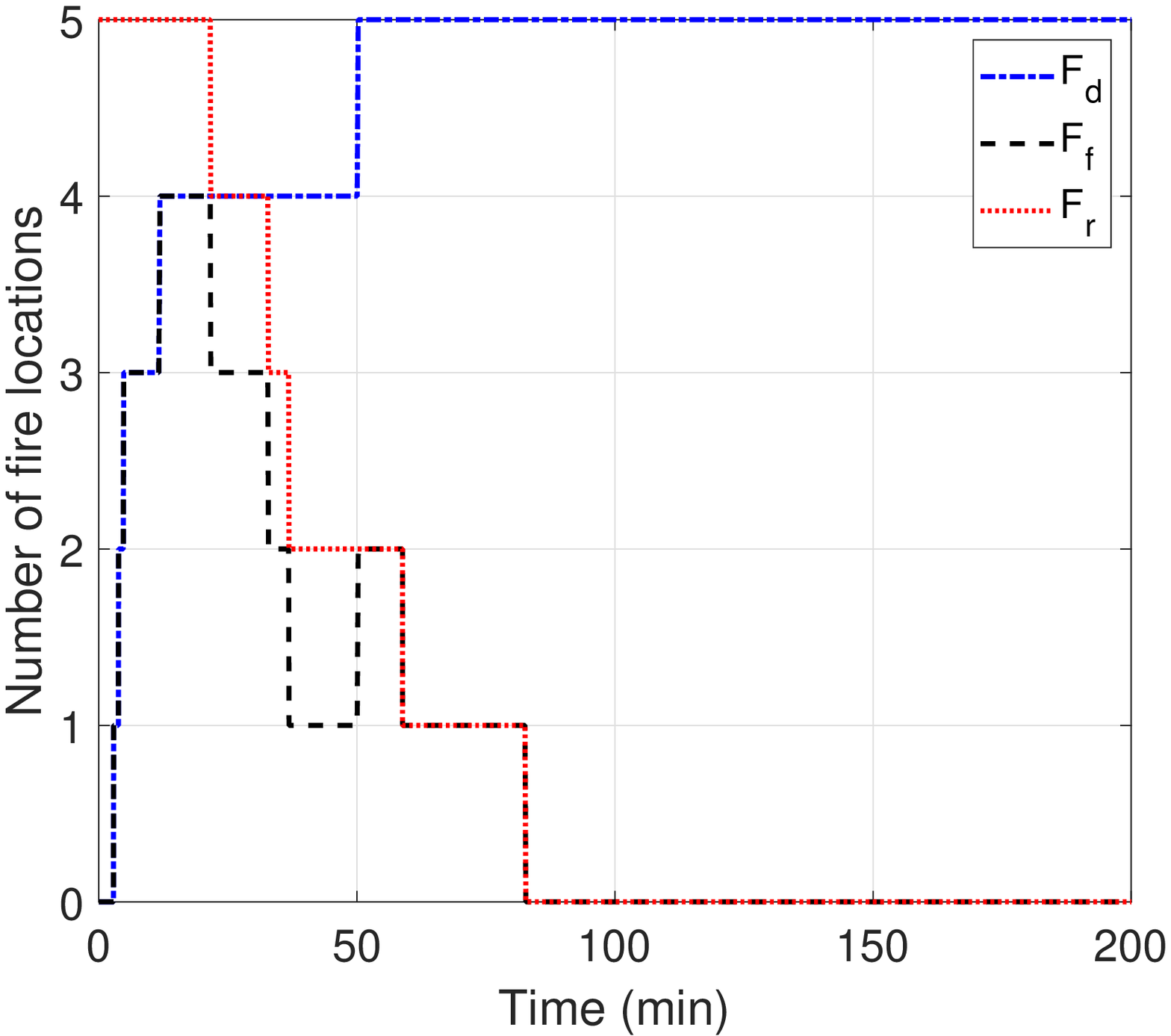}}%
		\subfloat[{} \label{fffdfr2}]
	{\includegraphics[trim=5 5 0 3,clip,width=4.5cm]{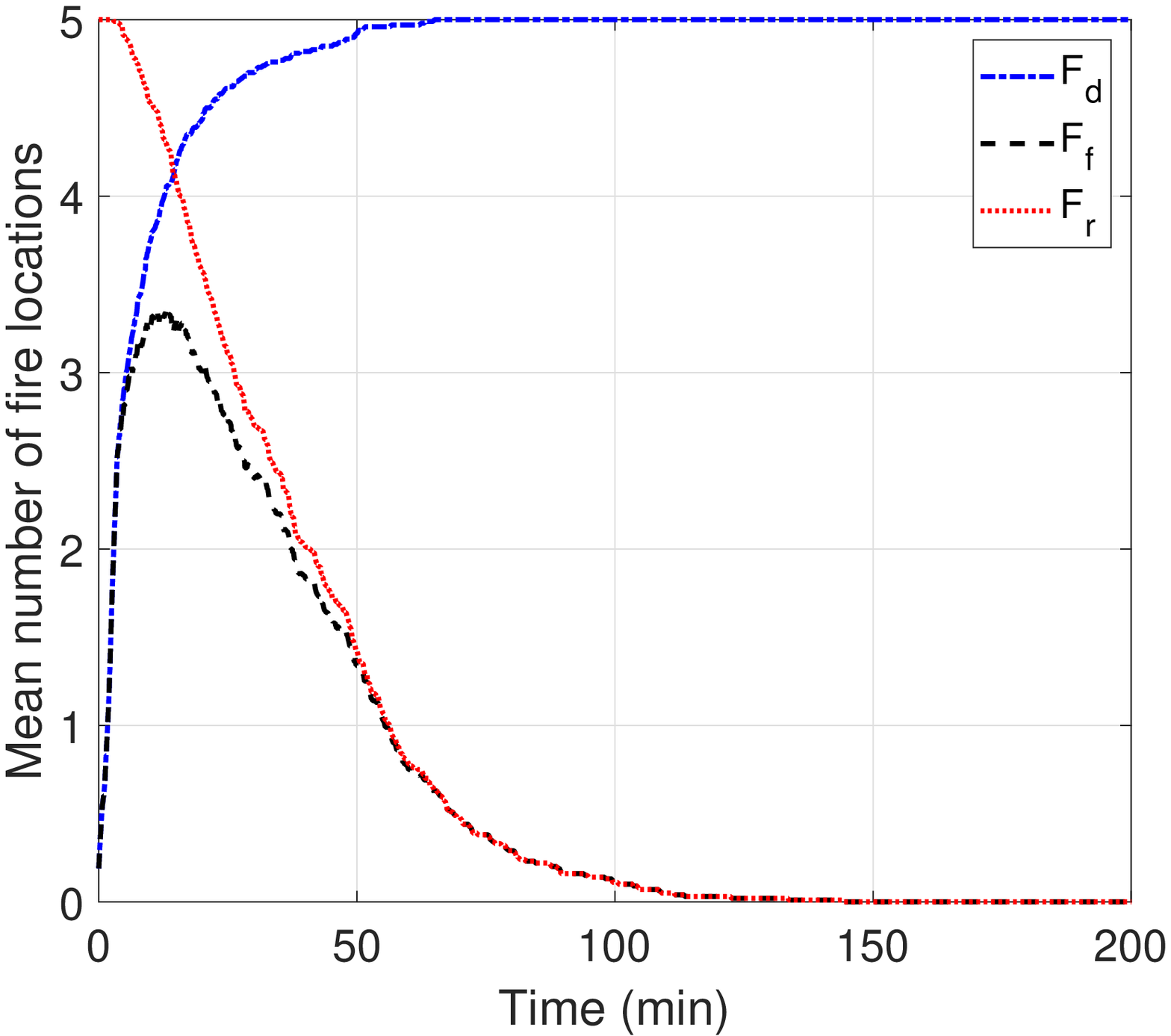}}%
	\caption{Number of fire locations detected ($F_d$), fire locations under mitigation ($F_f$), and fire locations remaining ($F_r $) for (a) single run (b) Monte-Carlo simulation.}
	\label{fffdfr}
	 \vspace{-3.5mm}
\end{figure}
The number of swarms executing the search ($S_s$) and quenching ($S_q$) for a total mission duration in $7$ swarm case is shown in Fig. \ref{SSSQ}. The number of swarms performing search decreases as more fire locations are detected and increases eventually when fires get quenched. About $50\%$ of the total mission time, swarms perform the search operation.

The number of fire locations detected ($F_d$ ), under mitigation ($F_f$), and remaining to be quenched ($F_r$), is shown in Fig. \ref{fffdfr}. It is necessary to maintain searching swarms to identify the unattended fires until all the fire locations are detected. The introduction of global regulative repulsion achieved this feature and the Monte-Carlo simulation results shown in Fig. \ref{SSSQ}b and \ref{fffdfr}b verify the feature.

\begin{table}[htbp]
\caption{Performance comparison for different methods}
	\centering
 \begin{tabular}{|c|c|c|c|c|}
\hline
\begin{tabular}[c]{@{}c@{}}Method \\ of forest\\firefighting\end{tabular} & \begin{tabular}[c]{@{}c@{}}Mean detection \\ time (min)\end{tabular} & \begin{tabular}[c]{@{}c@{}}Mean mission \\ time (min)\end{tabular} & Mean FER \\ \hline
Uniform-DFC & 66.10 & 217.47 & 1.808 \\\hline
Normal-DFC & 56.63 & 192.39 & 1.518 \\\hline
Levy-DFC & 52.59 & 184.14 & 1.398 \\\hline
OMS-DFC & 52.63 & 180.11 & 1.319 \\ \hline
MSCIDC & 24.13 & 71.56 & 0.477 \\ \hline
\end{tabular}
\label{diffdistr}
\end{table}

Table \ref{diffdistr} summarizes the mean detection time, mission time, and FER for different forest firefighting methods. Among the multi-UAV methods, OMS-DFC is superior with the lowest mean mission time and FER. Uniform search with DFC performs worst with the highest mean detection time, mission time and FER. The comparison study based on the Monte-Carlo simulation shows that the MSCIDC performs better with a $65\%$ reduction in the burned area and a $60\%$ reduction in mission time compared to OMS-DFC method.

\begin{figure}[htbp]
 \vspace{-5.5mm}
    \centering
	\subfloat[{} \label{area:a}]
	{\includegraphics[trim=5 5 0 5,clip,width=4.5cm]{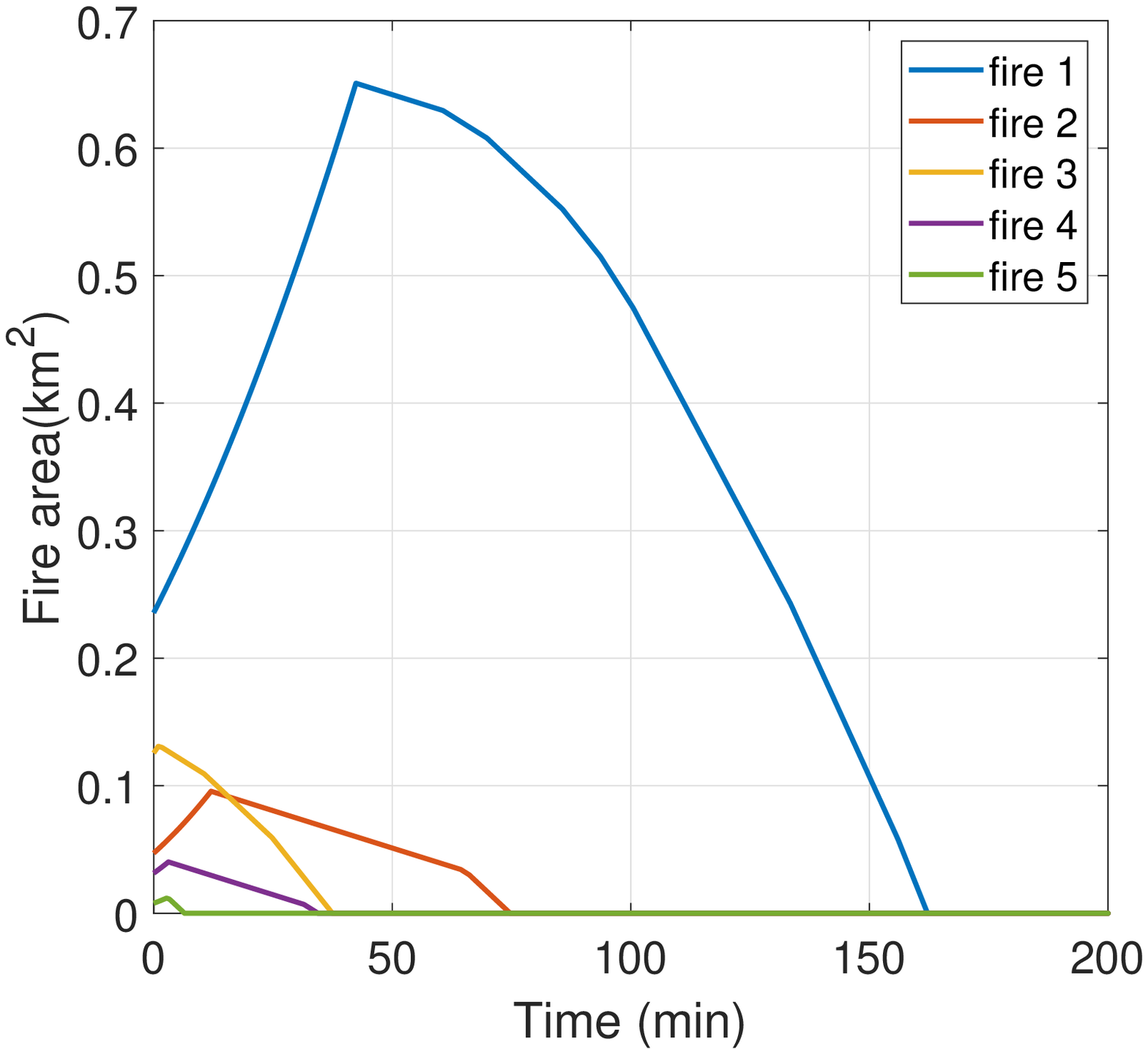}}%
		\subfloat[{} \label{area:b}]
	{\includegraphics[trim=5 5 0 5,clip,width=4.5cm]{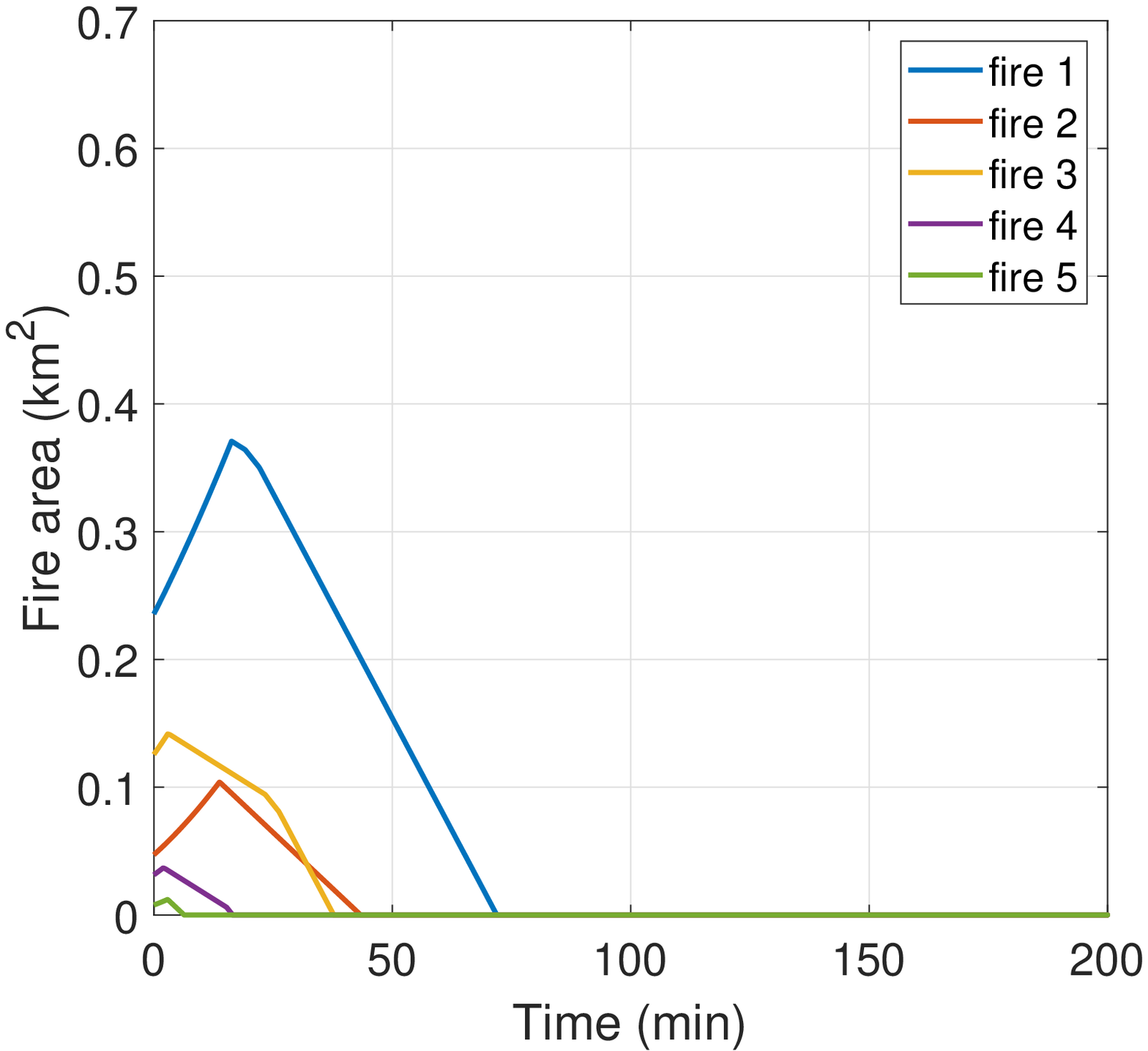}}%
	\caption{Comparison on variation of fire area with time for a single run between (a) OMS-DFC (b) MSCIDC.}
	\label{fireareaoms7swarm}
	\vspace{-2.5mm}
\end{figure}

	\begin{figure*}[t]
	\vspace{-5.5mm}
		\centering
		\subfloat[{} \label{main:a}]
		{\includegraphics[trim=15 2 50 20,clip,width=5.5cm]{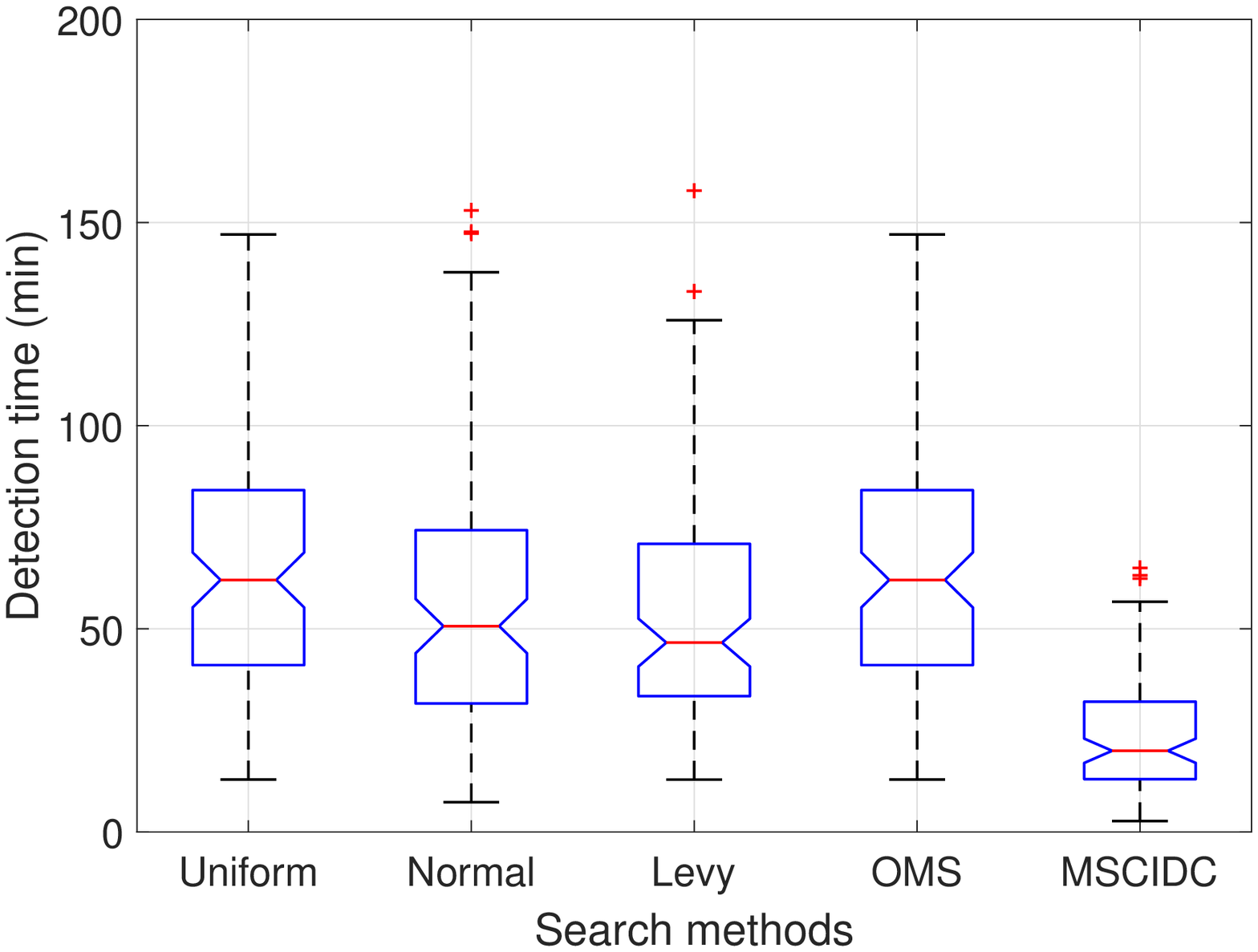}}%
		\subfloat[{} \label{main:b}] {      
			\includegraphics[trim=15 2 50 20,clip,width=5.5cm]{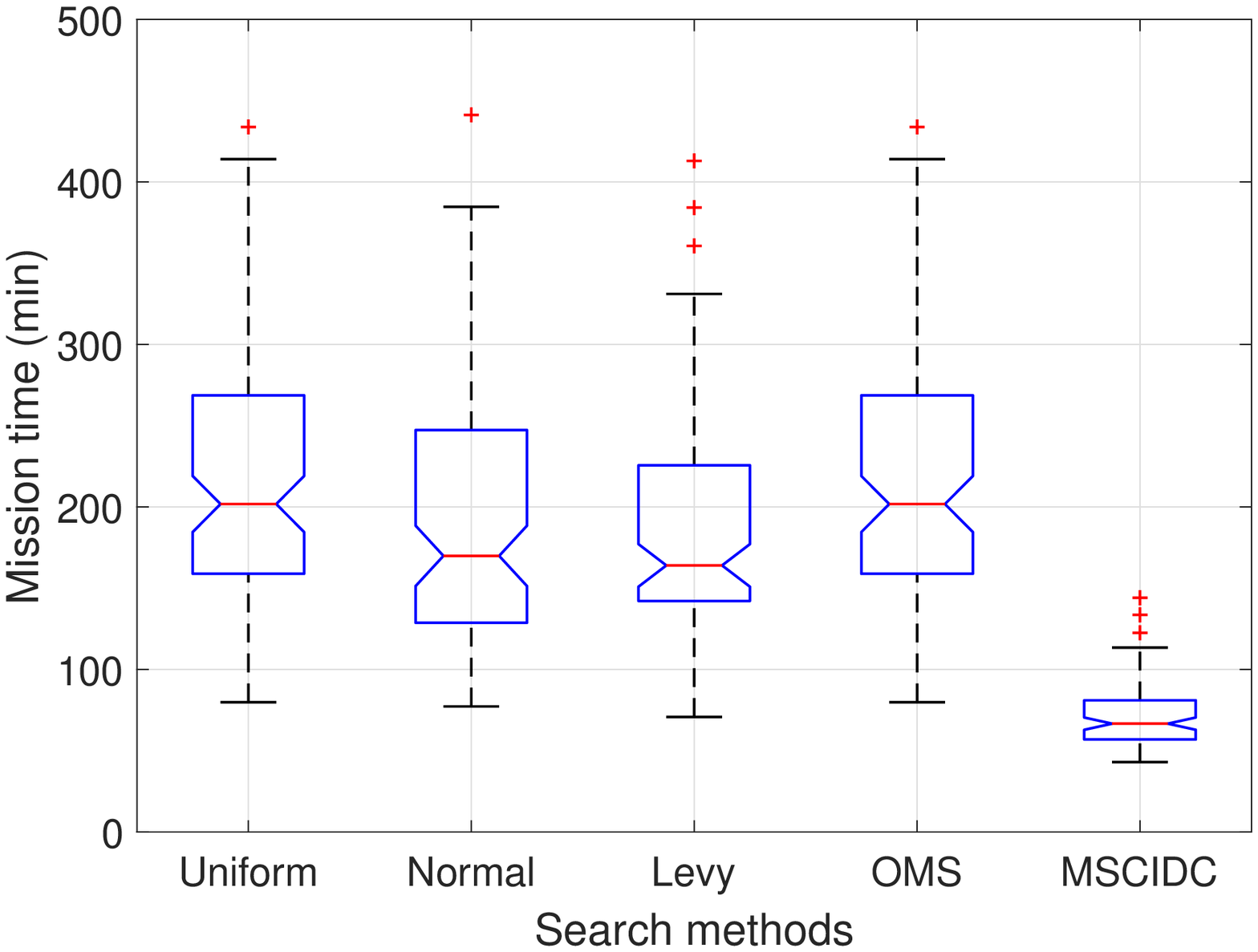}}%
		\subfloat[{}   \label{main:c}]{
			\includegraphics[trim=15 2 50 20,clip,width=5.5cm]{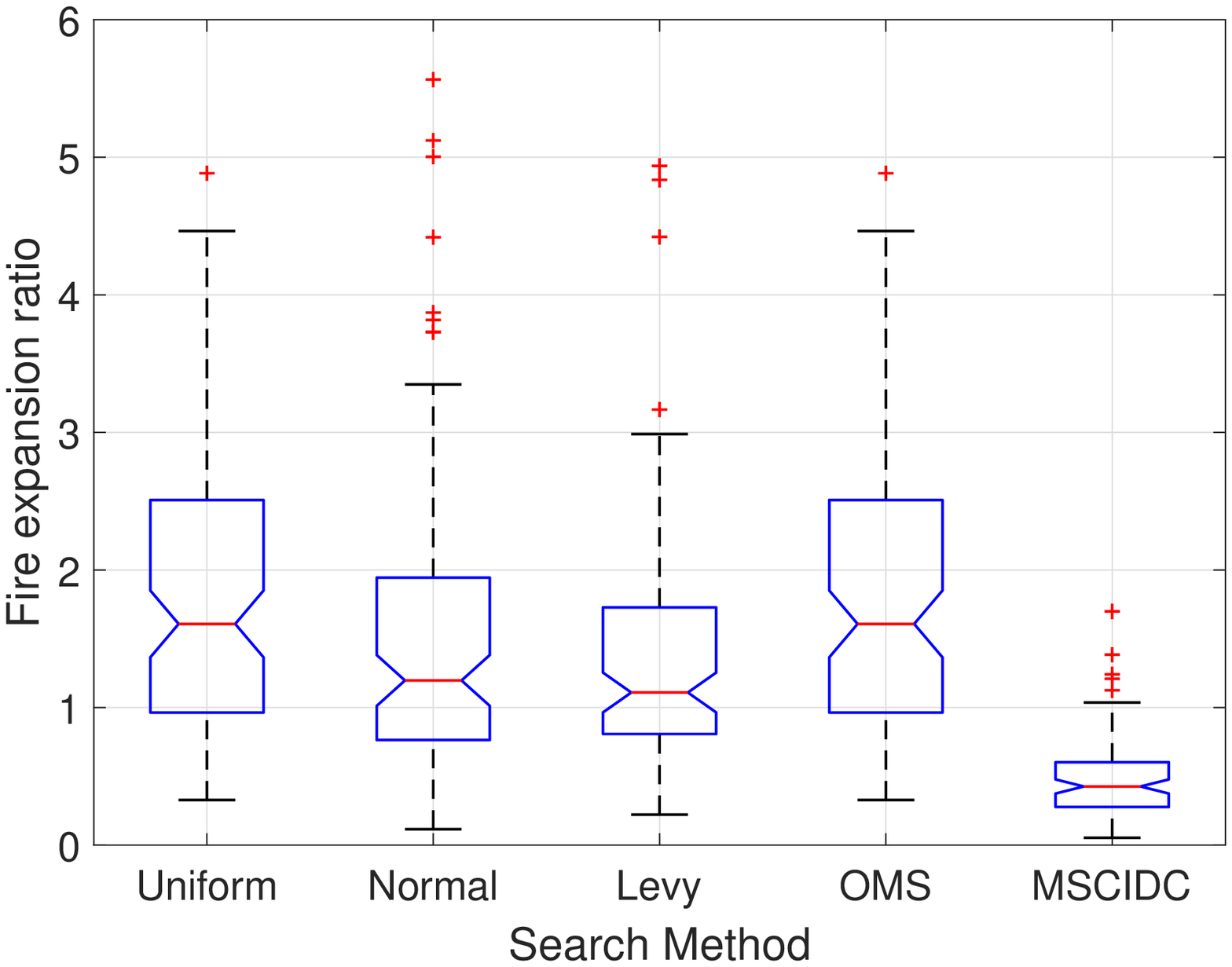}}%
		\caption{Box plot showing (a) detection time, (b) mission time and (c) fire expansion ratio for different methods}
		\label{comparison}
		\vspace{-5.5mm}
	\end{figure*}
The variation in the fire area of all fire locations versus time for a single run with OMS-DFC and MSCIDC is shown in Fig. \ref{fireareaoms7swarm}. The fire area increases until the swarms detect and start quenching the fire. For a single run shown in Fig. \ref{fireareaoms7swarm}, the burnt area accounts for $0.9295$ \si{km^2} with a FER of $1.07$ and $0.6654$ \si{km^2} with a FER of $0.48$ for OMS and MSCIDC, respectively. The fire spread area increases in the OMS method due to a higher detection time than the proposed method. The Fig. \ref{fireareaoms7swarm} also shows a faster mitigation as all swarm members detect the same fire location. The global regulative merging also accounts for reduced quench time.

Fig. \ref{comparison} shows the box plot of detection time, mission time, and FER. The box plots depict that the spread of the interquartile range of the proposed method is smaller than all the other existing methods. This indicates that existing multi-UAV methods have greater variability for performance indices compared to proposed method. The MSCIDC has lower search and mitigation time even though the UAVs are distributed and free to search in multi-UAV methods. The cooperative information sharing within the swarm help to detect target faster, and the swarm behavior leads to the detection of the same target by swarm members reducing the mitigation time compared to the multi-UAV cases. Even though there are outliers in plots of the proposed method, the outlier values are smaller than the $75^\text{th}$ percentile of the OMS method in all plots. The box plots of MSCIDC show a higher agreement to the median value of performance indices compared to multi-UAV methods.

\section{Conclusions}
This paper presents multi-swarm cooperative information-driven search and divide and conquer mitigation control for forest firefighting. The swarm members communicate the information sensed and cooperate in searching faster and efficiently. The members of the swarms are free to operate within the swarm boundary, and information sensed is used to decide the search direction and detect the fire locations. The local attraction between the swarm members maintains the swarm structure and helps swarm members reach the fire front faster. The divide and conquer mitigation control ensures that the areas quenched by UAVs are non-overlapped for effective and faster mitigation. The proposed method has no communication between the swarms during target search, and global regulative repulsion between swarms detecting the same targets benefits the detection of other unattended targets. The quenching time for higher capacity fires is effectively reduced with the global regulative merging of swarms. The Monte-Carlo analysis shows that the total time to detect all the five targets in the search area is less than $35\%$ of mission time. The proposed method is more efficient, with a reduction of more than $55\%$ in detection time and 60$\%$ mission time than the OMS method. The proposed method limits destruction of forest land area by $65\%$ compared to the OMS method. The results establish that the MSCIDC is more competent for diminishing the destruction of forest land by faster detection and mitigation of the fire location. The future work focuses on the allocation of firefighting tasks considering resource availability and target capacity. 

\bibliographystyle{IEEEtran}
\balance
\bibliography{forestfire}

\vfill

\end{document}